\documentclass{aa}
\usepackage{rotating}
\usepackage{graphicx}
\usepackage{txfonts}
\usepackage{natbib}
\begin{document}

\title{Fundamental parameters of massive stars in multiple systems: The cases of HD~17505A and HD~206267A\thanks{Based on observations collected with the TIGRE telescope (La Luz, Mexico), the 1.93~m telescope at Observatoire de Haute Provence (France), the Nordic Optical Telescope at the Observatorio del Roque de los Muchachos (La Palma, Spain), and the Canada-France-Hawaii telescope (Mauna Kea, Hawaii).}}
\author{F.\ Raucq\inst{1} \and G.\ Rauw\inst{1} \and L.\ Mahy\inst{2,1}\fnmsep\thanks{FRS-FNRS Postdoctoral Researcher} \and S.\ Sim\'on-D\'{\i}az\inst{3,4}}

\institute{Space sciences, Technologies and Astrophysics Research (STAR) Institute, Universit\'e de Li\`ege, All\'ee du 6 Ao\^ut, 19c, B\^at B5c, 4000 Li\`ege, Belgium
\and Instituut voor Sterrenkunde, KU Leuven, Celestijnenlaan 200D, Bus 2401, 3001 Leuven, Belgium
\and Instituto de Astrof\'{\i}sica de Canarias, E-38200 La Laguna, Tenerife, Spain \and Departamento de Astrof\'{\i}sica, Universidad de La Laguna, E-38205 La Laguna, Tenerife, Spain}

\authorrunning{Raucq et al.}
\titlerunning{HD~17505 and HD~206267}
\offprints{G.\ Rauw}
\mail{rauw@astro.ulg.ac.be}
\date{}
\keywords{Stars: early-type -- binaries: spectroscopic -- Stars: fundamental parameters -- Stars: massive -- Stars: individual: HD~17505A -- Stars: individual: HD~206267A}


\abstract{Many massive stars are part of binary or higher multiplicity systems. The present work focusses on two higher multiplicity systems: HD~17505A and HD~206267A.}{Determining the fundamental parameters of the components of the inner binary of these systems is mandatory to quantify the impact of binary or triple interactions on their evolution.}{We analysed high-resolution optical spectra to determine new orbital solutions of the inner binary systems. After subtracting the spectrum of the tertiary component, a spectral disentangling code was applied to reconstruct the individual spectra of the primary and secondary. We then analysed these spectra with the non-LTE model atmosphere code CMFGEN to establish the stellar parameters and the CNO abundances of these stars.}{The inner binaries of these systems have eccentric orbits with $e \sim 0.13$ despite their relatively short orbital periods of 8.6 and 3.7\,days for HD~17505Aa and HD~206267Aa, respectively. Slight modifications of the CNO abundances are found in both components of each system. The components of HD~17505Aa are both well inside their Roche lobe, whilst the primary of HD~206267Aa nearly fills its Roche lobe around periastron passage. Whilst the rotation of the primary of HD~206267Aa is in pseudo-synchronization with the orbital motion, the secondary displays a rotation rate that is higher.}{The CNO abundances and properties of HD~17505Aa can be explained by single star evolutionary models accounting for the effects of rotation, suggesting that this system has not yet experienced binary interaction. The properties of HD~206267Aa suggest that some intermittent binary interaction might have taken place during periastron passages, but is apparently not operating anymore.}

\maketitle

\section{Introduction}
Binary systems are important tools for observationally determining the masses and radii of stars. However, the binarity also implies far more complex and diverse evolutionary paths for these systems \citep[e.g.][]{Vanbeveren17,Langer08,Langer07}. This is especially relevant for massive stars since the incidence of binary or higher multiplicity systems is high among these objects \citep[][and references therein]{Duchene13}. In close binaries, the exchange of mass and angular momentum between the stars leads to various observational signatures such as over- or under-luminosities, peculiar chemical abundances, and asynchronous rotation.\ \citep[e.g.][]{Linder08,Mahy11,Raucq16,Raucq17}. The evolution of the inner pair of stars in gravitationally-bound hierarchical triple systems is even more complex, since the Lidov-Kozai cycles can modulate the eccentricity of the inner binary, leading to modulations of the binary interactions \citep{Toonen16}. Observational constraints on these phenomena can be obtained through in-depth analyses of the spectra of such systems using spectral disentangling methods coupled to modern model atmosphere codes \citep[e.g.][]{Raucq16,Raucq17}. In this paper, we apply this approach to the spectra of two close massive binaries, HD~17505Aa and HD~206267Aa, which are part of higher multiplicity systems. 

HD~17505 is an O-star system dominating the centre of the cluster IC~1848 within the Cas~OB6 stellar association at a distance of about 2.3\,kpc \citep{Garmany92,Massey95}, which was revised to 1.9\,kpc by \citet{Hillwig06}. Previous studies suggested that HD~17505 is a multiple system consisting of at least four O stars (with a total mass approaching 100\,M$_{\odot}$) that are apparently gravitationally bound \citep[][and references therein]{Stickland01,Hillwig06}, and up to six visual companions that have not been shown to be gravitationally bound \citep{Mason98}, at angular separations ranging from 2\arcsec to 124\arcsec. The inner binary \citep[HD~17505Aa; see][]{Hillwig06} is composed of two O7.5\,V stars with a short orbital period of 8.57\,days. The inner binary spectra are blended with that of a third component (HD~17505Ab), which is classified as an O6.5\,III star with an orbital period of less than 61\,years \citep{Hillwig06}. 

HD~206267 is a trapezium-like system \citep[e.g.][]{Mason13} dominating the young open cluster Trumpler~37 and the H\,{\sc ii} region IC~1396, embedded in the Cep~OB2 association. The system was first studied by \citet{Plaskett23}, who noted that its spectrum displays rather diffuse lines, some of which were double and showed variable strength and profile. \citet{Stickland95} showed the main component of the system (HD~206267A) to be triple, and presented an orbital solution of the inner binary, HD\,206267Aa, based on high-resolution, short-wavelength {\it IUE} spectra. He derived an orbital period of 3.71\,days for the inner binary system. The third component, HD~206267Ab, was found to display a constant velocity. \citet{Burkholder97} subsequently revised the orbital solution of the inner binary and inferred spectral types of O6.5\,V and O9.5\,V for the primary and secondary, respectively.

This paper is organized as follows. Section \ref{obs} describes the spectroscopic observations and data reduction. Revised orbital solutions are derived in Sect.\,\ref{New-orbital-solutions}. Section \ref{preparation} presents the preparatory work on the sample of spectra aiming at the reconstruction of the individiual spectra of the components of both binary systems. The resulting spectra are then used in Sect.\,\ref{analysis} to derive the fundamental stellar parameters, notably through fits with a model atmosphere code. Finally, Sect.\,\ref{summary} summarizes our main results and discusses their implications.

\section{Observations and data reduction \label{obs}}
Fifteen high-resolution optical spectra of HD~17505A and 22 spectra of HD~206267A were obtained with the HEROS spectrograph, mounted on the 1.2\,m TIGRE telescope at La Luz Observatory \citep[Mexico;][]{Schmitt14}. The HEROS spectrograph covers two spectral domains, ranging from 3500\,\AA\ to 5600\,\AA\ (the blue channel) and from 5800\,\AA\ to 8800\,\AA\ (red channel), with a resolution of $\sim$20\,000. The spectra were reduced with an Interactive Data Language (IDL) pipeline \citep{Schmitt14} based on the reduction package REDUCE written by \citet{Piskunov02}. The spectra of HD~17505A had a signal-to-noise ratio (S/N) of $\sim 100$ between 4560\,\AA\ and 4680\,\AA, whilst this number was $\sim 150$ for HD~206267A.

We complemented the HEROS spectra with a series of data from various archives. One reduced spectrum of HD~17505A and one of HD~206267A were extracted from the ELODIE archive. The ELODIE {\'e}chelle spectrograph was mounted on the 1.93\,m telescope of the Observatoire de Haute Provence (OHP, France) between 1993 and 2006. It covered the spectral range from 3850\,\AA\ to 6800\,\AA\ with a resolution of $\sim$42\,000. Six spectra of HD~17505A and five of HD~206267A were obtained as part of the IACOB project \citep{SimonDiaz11a,SimonDiaz11b,SimonDiaz15} with the FIES {\'e}chelle spectrograph. The FIES instrument is mounted on the 2.5\,m Nordic Optical Telescope located at the Observatorio del Roque de los Muchachos (La Palma, Spain). This spectrograph covers the spectral range from 3700\,\AA\ to 7300\,\AA\ with a resolving power of $\sim$46\,000 in medium-resolution mode.

For HD~206267A, we further took one spectrum from the ESPaDOnS archives and nine spectra from the SOPHIE archives. The ESPaDOnS {\'e}chelle spectropolarimeter has a resolving power of $\sim 68\,000$ over the complete optical spectrum and is operated on the 3.6\,m Canada-France-Hawaii Telescope on Mauna Kea, whereas the SOPHIE spectrograph is mounted on the 1.93\,m telescope at OHP, and covers the wavelength range from 3872\,\AA\ to 6943\,\AA\ at a resolution of $\sim$40 000 (high-efficiency mode). 

Since our analysis combines spectra collected with different instruments, one might wonder about systematic differences in the wavelength calibration of the spectra. To quantify this, we measured the velocities of the interstellar Na\,{\sc i} D$_1$ and D$_2$ lines for those instruments where more than one spectrum had been taken. The highest resolution spectra (FIES, SOPHIE) reveal that these lines actually consist of several components that are heavily blended. Yet, since they are not resolved on the HEROS data, which make up the bulk of our data, we simply fitted a single Gaussian to each line. The results of these tests are shown in Table\,\ref{wavelength}. Within the errors, the velocities determined with the different instruments overlap with the mean value determined from all data. Systematic differences are found to be less than 2\,km\,s$^{-1}$, and part of these differences are most likely due to the non-Gaussian shape of the lines seen with the different resolving powers. These systematic differences are much smaller than the typical errors on the measurements of the stellar radial velocities (RVs). The journals of the observations of HD~17505A and HD~206267A are presented in Tables\,\ref{journal_17505} and \ref{journal_206267}, respectively.
\begin{table}
\caption{Heliocentric radial velocities of the interstellar Na\,{\sc i} lines.\label{wavelength}}
\begin{center} 
\begin{tabular}{c | c c | c c}
\hline
& \multicolumn{2}{c}{HD~17505A} & \multicolumn{2}{c}{HD~206267A}\\
\cline{2-5}
 & Na\,{\sc i} D$_1$ & Na\,{\sc i} D$_2$ & Na\,{\sc i} D$_1$ & Na\,{\sc i} D$_2$\\
 & (km\,s$^{-1}$) & (km\,s$^{-1}$) & (km\,s$^{-1}$) & (km\,s$^{-1}$) \\
\hline
FIES   & $-15.0 \pm 1.4$ & $-16.5 \pm 1.5$ & $-13.3 \pm 1.2$ & $-15.0 \pm 1.3$ \\ 
SOPHIE & -- & -- & $-13.6 \pm 1.5$ & $-15.7 \pm 0.1$ \\
HEROS  & $-13.0 \pm 1.6$ & $-14.9 \pm 1.6$ & $-11.1 \pm 0.6$ & $-13.3 \pm 0.8$ \\
\hline
All data & $-13.5 \pm 1.8$ & $-15.4 \pm 1.7$ & $-12.0 \pm 1.5$ & $-14.1 \pm 1.3$\\
\hline
\end{tabular}
\end{center}
\end{table}

The vast majority of the spectra listed in Tables \ref{journal_17505} and \ref{journal_206267} have very similar S/N ratios. For the spectral disentangling, we thus considered all data with equal weights. The spectra were continuum normalized using a spline-fitting method under the Munich Image Data Analysis System ({\tt MIDAS}) software. For each system, we adopted a single set of carefully chosen continuum windows to normalize all spectra in a self-consistent way. Finally, for the purpose of spectral disentangling, all spectra were rebinned with a wavelength step of 0.02\,\AA.

\begin{table*}
\caption{Journal of the observations of HD~17505A.\label{journal_17505}}
\begin{center}
\begin{tabular}{c|ccc|ccc}
\hline 
HJD $-2\,450\,000$ & Instrument & Exp.\ time & $\phi$ & RV(Aa1) & RV(Aa2) & RV(Ab)\\
&  & (min) &  & \multicolumn{3}{c}{(km\,s$^{-1}$)}\\
\hline 
3327.473 & ELODIE & 60 & 0.75 & -191.0 & 114.3 & -31.3\\
5447.762 & FIES & 15 & 0.19 & 115.5 & -193.3 & -31.7\\
5577.414 & FIES & 13.7 & 0.32 & 90.8 & -163.3 & -1.15:\\
5812.692 & FIES & 11.8 & 0.78 & -208.3 & 113.0 & -13.6:\\
5814.677 & FIES & 9.2 & 0.01 & -28.1 & -28.1 & -13.6:\\
5816.755 & FIES & 8.3 & 0.25 & 106.3 & -200.4 & -13.6:\\
6285.436 & FIES & 15 & 0.95 & -128.1 & 43.0 & 9.1:\\
6895.965 & HEROS & 60 & 0.20 & 126.2 & -160.5 & -37.6\\
6897.853 & HEROS & 60 & 0.42 & 36.3: & -88.5: & -37.6\\
6907.940 & HEROS & 60 & 0.60 & -36.1 & -4.1: & -37.6\\
6910.888 & HEROS & 40 & 0.94 & -136.1 & 80.5 & -37.6\\
6911.915 & HEROS & 60 & 0.06 & -38.2 & -38.2 & -37.6\\
6918.869 & HEROS & 60 & 0.88 & -168.5 & 126.3 & -37.6\\
6925.891 & HEROS & 60 & 0.69 & -127.1: & 94.7 & -37.6\\
6939.829 & HEROS & 60 & 0.32 & 139.1 & -152.3 & -37.6\\
6941.814 & HEROS & 60 & 0.55 & -28.2 & -28.2 & -37.6\\
6943.801 & HEROS & 60 & 0.79 & -171.4 & 136.1 & -37.6\\
6945.770 & HEROS & 60 & 0.01 & -26.4 & -26.4 & -37.6\\
6947.861 & HEROS & 60 & 0.26 & 144.8 & -166.1 & -37.6\\
6949.855 & HEROS & 20 & 0.49 & -21.8 & -21.8 & -37.6\\
6953.787 & HEROS & 20 & 0.95 & -126.8 & 85.6 & -37.6\\
6955.785 & HEROS & 60 & 0.18 & 128.8 & -142.9 & -37.6\\
\hline 
\end{tabular}
\end{center}
\tablefoot{The phases ($\phi$) are computed according to the ephemerides of the new orbital solution (labelled ``This study'' in Table\,\ref{Orb_sol_17505_tab}). 
The typical uncertainties on the RVs are $5-15$ km~s$^{-1}$. The colons indicate uncertainties larger than 20 km~s$^{-1}$. For the third component, we calculated the mean value of the RVs over each run of observations, given its long orbital period.}
\end{table*}

\begin{table*}
\caption{Journal of the observations of HD~206267A.\label{journal_206267}}
\begin{center}
\begin{tabular}{c|ccc|cc}
\hline 
HJD$-2\,450\,000$ & Instrument & Exp.\ time & $\phi$ & RV(Aa1) & RV(Aa2)\\
 &  & (min) &  & (km\,s$^{-1}$) & (km\,s$^{-1}$)\\
\hline 
0710.4441 & ELODIE & 22.3 & 0.42 & -166.64: & 279.78\\
5494.3031 & FIES & 3.3 & 0.95 & 163.71 & -298.86\\
5729.1077 & ESPaDOnS & 7.5 & 0.24 & -144.38 & 95.92\\
5812.5828 & FIES & 2.1 & 0.74 & 107.95 & -113.76:\\
5816.4969 & FIES & 2.1 & 0.80 & 116.34: & -141.13:\\
5817.5170 & FIES & 2.6 & 0.07 & 70.20: & -233.19\\
6285.3231 & FIES & 3.4 & 0.17 & -7.10 & -7.10\\
6527.4539 & SOPHIE & 5 & 0.44 & -160.71 & 294.96\\
6527.5577 & SOPHIE & 5 & 0.47 & -160.47 & 291.80\\
6528.4396 & SOPHIE & 5 & 0.71 & -13.17 & -13.17\\
6528.5550 & SOPHIE & 5 & 0.74 & -13.17 & -13.17\\
6529.4943 & SOPHIE & 5 & 0.99 & 180.19 & -301.36\\
6530.3561 & SOPHIE & 5 & 0.22 & -137.16 & 80.18\\
6530.5523 & SOPHIE & 5 & 0.28 & -149.83 & 164.45\\
6531.4081 & SOPHIE & 5 & 0.51 & -170.80 & 251.79\\
6531.5479 & SOPHIE & 5 & 0.54 & -140.55 & 221.30\\
6895.7500 & HEROS & 30 & 0.72 & -4.52 & -4.52\\
6897.7738 & HEROS & 30 & 0.26 & -133.61 & 128.00\\
6907.7219 & HEROS & 30 & 0.95 & 123.25: & -312.95:\\
6909.7156 & HEROS & 30 & 0.48 & -147.36 & 291.91\\
6911.7011 & HEROS & 30 & 0.02 & 149.51 & -279.74\\
6920.6768 & HEROS & 30 & 0.44 & -162.35 & 276.12:\\
6926.6468 & HEROS & 30 & 0.05 & 159.82 & -252.97\\
6939.6036 & HEROS & 30 & 0.54 & -145.36 & 223.14\\
6941.6020 & HEROS & 30 & 0.08 & 167.72 & -227.50\\
6944.6560 & HEROS & 30 & 0.90 & 160.05 & -278.83\\
6957.5806 & HEROS & 30 & 0.38 & -146.99 & 273.99:\\
7236.7624 & HEROS & 15 & 0.64 & -142.53 & 164.15\\
7236.7729 & HEROS & 15 & 0.64 & -117.50 & 165.15\\
7238.7444 & HEROS & 30 & 0.17 & -3.96 & -3.96\\
7264.8159 & HEROS & 30 & 0.20 & -3.96 & -3.96\\
7271.7659 & HEROS & 30 & 0.08 & 161.57 & -229.72\\
7286.6714 & HEROS & 30 & 0.09 & 162.89 & -180.75\\
7288.6617 & HEROS & 30 & 0.63 & -145.56 & 131.03:\\
7290.7237 & HEROS & 30 & 0.19 & -3.96 & -3.96\\
7292.6865 & HEROS & 30 & 0.72 & -92.04: & 158.66\\
7308.6686 & HEROS & 30 & 0.02 & 170.45 & -278.45\\
7323.5769 & HEROS & 30 & 0.04 & 178.21 & -257.39\\
\hline 
\end{tabular}
\end{center}
\tablefoot{The phases ($\phi$) are computed according to the ephemerides of the combined orbital solution listed in Table\,\ref{Orb_sol_206267}. 
The typical uncertainties on the RVs are $10-20$ km~s$^{-1}$. The colons indicate uncertainties larger than 30 km~s$^{-1}$. The mean value of RV(Ab) is $-7.0 \pm 7.8$ km\,s$^{-1}$.}
\end{table*}

\section{New orbital solutions\label{New-orbital-solutions}}
\subsection{HD~17505Aa}
Using our set of observations, we revised the orbital solution of the inner binary system of HD~17505A. To do so, we concentrated our efforts on the strongest absorption lines that are essentially free from blends with other features. In this way, we measured the RVs of the H$\gamma$, He\,{\sc i} $\lambda\lambda$\,4471, 5876, 7065, He\,{\sc i} + {\sc ii} $\lambda$\,4026, and He\,{\sc ii} $\lambda\lambda$\,4542, 5412 lines via a multi-Gaussian fit. We adopted the effective wavelengths from \citet{Underhill95} and \citet{Ninkov}. The primary and secondary stars of the inner binary, HD~17505Aa, apparently display the same spectral type, which leads to some difficulties in distinguishing the two stars. For each observation, the RVs of the primary and secondary stars quoted in Table\,\ref{journal_17505} were determined as the mean of the corresponding RVs measured for each of the above-listed lines in that given observation. Since all our data points were obtained with the same set of lines, we obtain a coherent set of RVs that should be free of any bias due to the choice of the lines under consideration\footnote{We note that such biases should be rather small for binary systems consisting of detached main-sequence stars \citep{Palate}.}. For most data points, the uncertainties on the RVs are about 5 -- 15\,km~s$^{-1}$. In some cases (indicated by the colons in Table\,\ref{journal_17505}), the uncertainties however exceed 20\,km~s$^{-1}$.

We then performed a Fourier analysis of the RV(Aa1) and RV(Aa2) data using the Fourier-method for uneven sampling of \citet{HMM}, modified by \citet{Gosset}. This analysis yielded the highest peak around $\nu=0.11670 \pm 0.00003$\,d$^{-1}$, i.e.\ $P_{\rm orb}=8.5690 \pm 0.0022$\,days, which is consistent within the error bars with the orbital period determined in the study of \citet{Hillwig06}, $P_{\rm orb} = 8.5710 \pm 0.0008$\,days. For completeness, we also performed the same Fourier analysis with our RVs combined to those reported by \citet{Hillwig06} and found again the highest peak at $\nu=0.11670 \pm 0.00002$\,d$^{-1}$, indicating that our period determination $P_{\rm orb}=8.5690 \pm 0.0014$\,days is consistent with their work.

\begin{figure*}
\begin{center}
\begin{minipage}{8.5cm}
\resizebox{8.5cm}{!}{\includegraphics{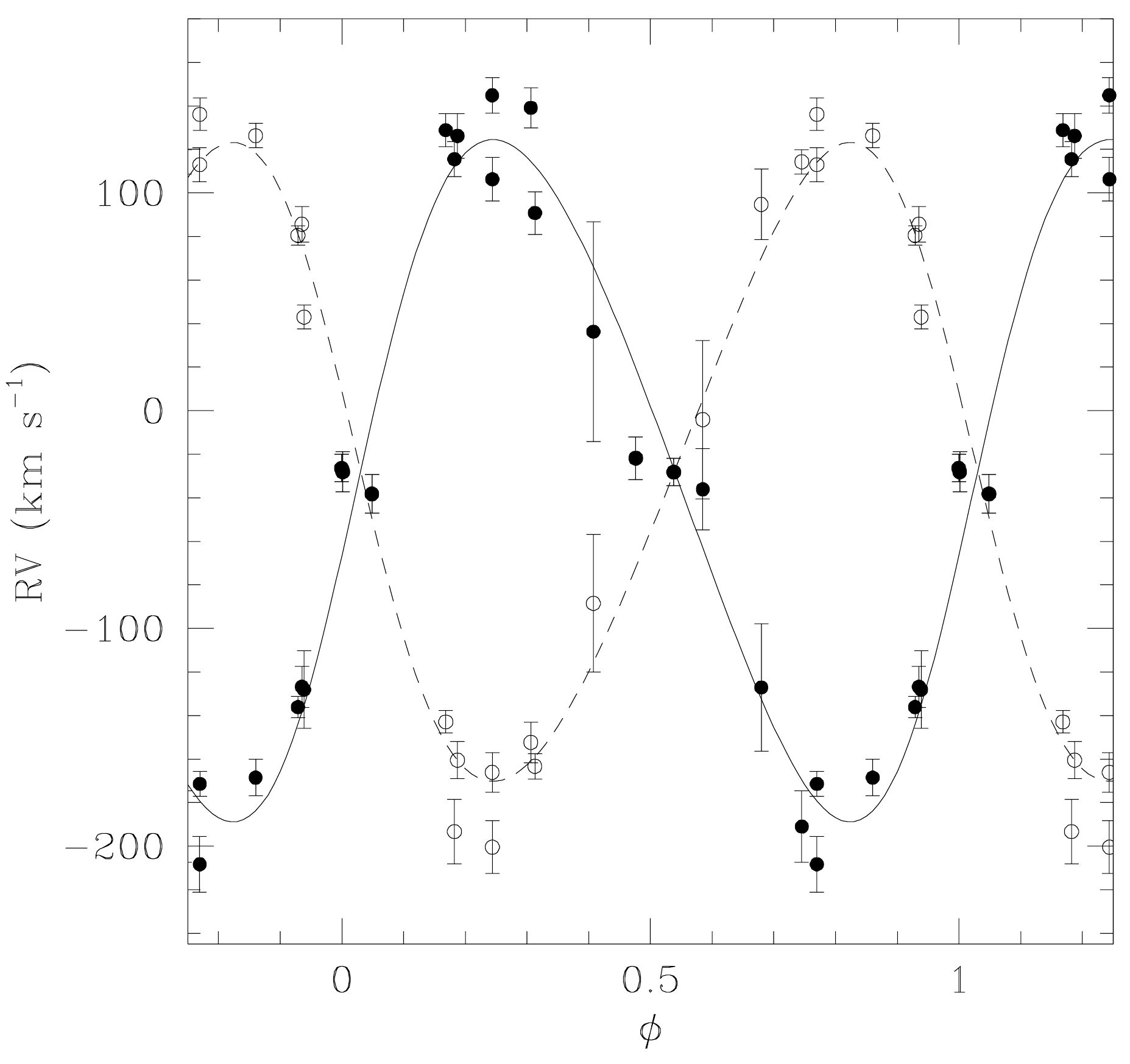}}
\end{minipage}
\hfill
\begin{minipage}{8.5cm}
\resizebox{8.5cm}{!}{\includegraphics{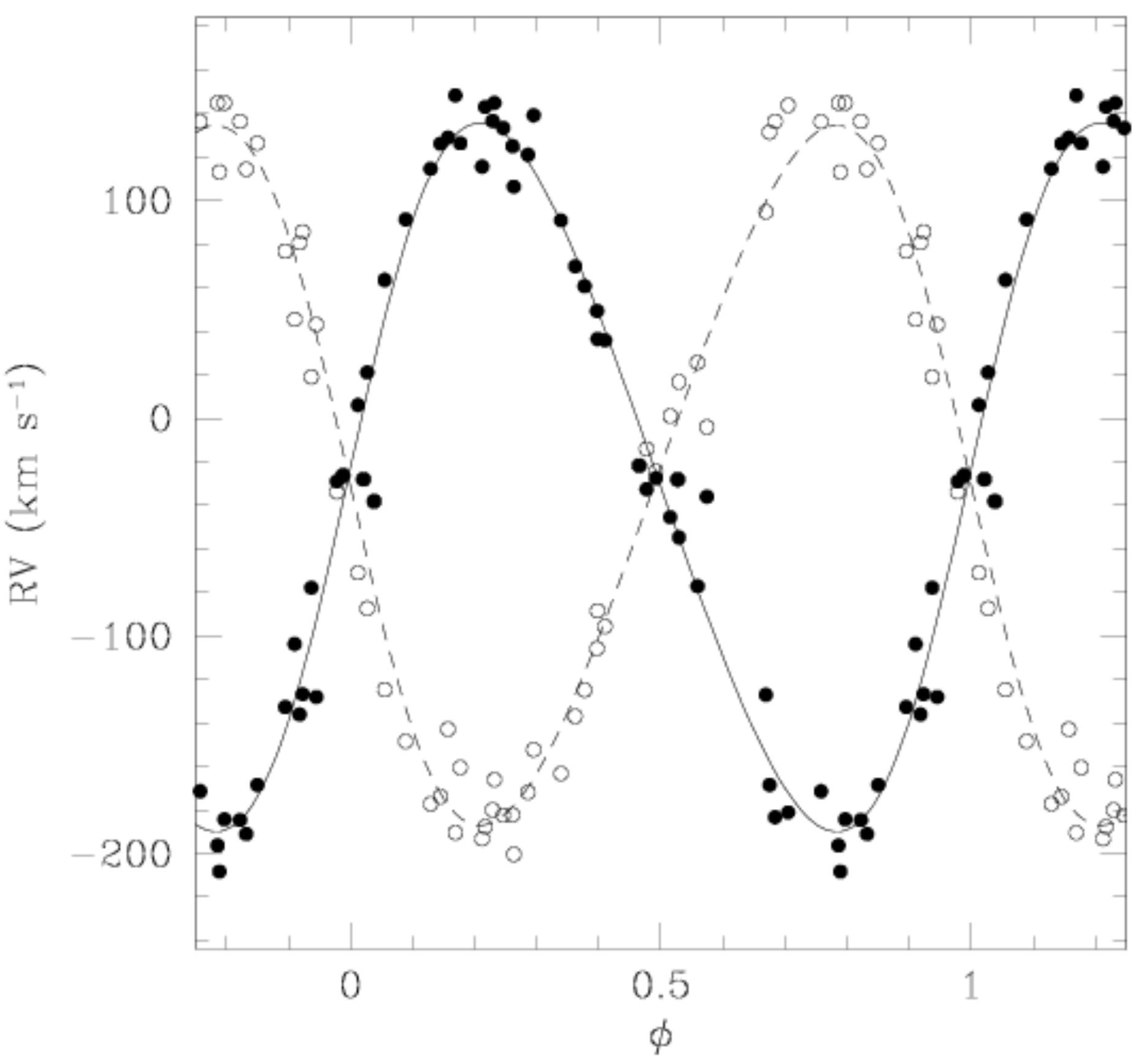}}
\end{minipage}
\end{center}
\caption{Radial velocities of HD~17505Aa1 (filled circles) and HD~17505Aa2 (open circles), assuming a period of 8.5690\,days. The left panel shows the RVs corresponding to the new measurements (Table\,\ref{journal_17505}), while the right panel shows the solution obtained by combining our measurements with those of \citet{Hillwig06}. The solid and dashed lines indicate the orbital solutions from Table\,\ref{Orb_sol_17505_tab}. \label{orb_sol_fig_17505}}
\end{figure*}

Adopting an orbital period of 8.5690\,days, we computed an orbital solution with the Li\`ege Orbital Solution Package code \citep[LOSP,][and references therein]{Sana09}. The RVs were weighted according to their estimated uncertainties. The result is shown in the left panel of Fig.\,\ref{orb_sol_fig_17505} and the corresponding orbital elements are provided in Table\,\ref{Orb_sol_17505_tab}. We further computed an orbital solution combining our newly measured RVs with those of \citet{Hillwig06}. The RV curve is shown in the right panel of Fig.\,\ref{orb_sol_fig_17505} and the corresponding orbital elements are also given in Table\,\ref{Orb_sol_17505_tab}.

The RV amplitudes of both our single and combined solutions are lower than reported by \citet{Hillwig06}, and our determined eccentricities are slightly larger than reported by these authors (see Table\,\ref{Orb_sol_17505_tab}). The lower RV amplitudes result in lower minimum masses compared to the solution of \citet{Hillwig06}.

\begin{table*}
\caption{Orbital solution of HD~17505Aa computed from our RV data, assuming an eccentric orbit and a period of 8.5690\,days, compared to the orbital solution proposed by \citet{Hillwig06} and an orbital solution obtained by combining our RVs with those measured by these authors.\label{Orb_sol_17505_tab}}
\begin{center}
\begin{tabular}{l|cc|cc|cc}
\hline 
 & \multicolumn{2}{c|}{This study} & \multicolumn{2}{c|}{\citet{Hillwig06}} & \multicolumn{2}{c}{Combined solution}\\
 & Prim.\ (Aa1) & Seco.\ (Aa2) & Prim.\ (Aa1) & Seco.\ (Aa2) & Prim.\ (Aa1) & Seco.\ (Aa2)\\
\hline 
$P_{\rm orb}$ (days) & \multicolumn{2}{c|}{$8.5690 \pm 0.0022$} & \multicolumn{2}{c|}{$8.5710 \pm 0.0008$} & \multicolumn{2}{c}{$8.5690 \pm 0.0014$} \\ 
$T_0$ (HJD $- 2\,450\,000$) & \multicolumn{2}{c|}{3329.656 $\pm$ 0.313} & \multicolumn{2}{c|}{1862.696 $\pm$ 0.016} & \multicolumn{2}{c}{3328.910 $\pm$ 0.173}\\
$e$ & \multicolumn{2}{c|}{0.128$\pm$ 0.037} & \multicolumn{2}{c|}{0.095$\pm$ 0.011} & \multicolumn{2}{c}{0.118$\pm$ 0.018}\\
$\gamma$ (km\,s$^{-1}$) & $-27.8 \pm 4.7$ & $-27.6 \pm 4.5$ & $-25.8 \pm 1.8$ & $-26.3 \pm 1.2$ & $-27.7 \pm 2.3$ & $-25.9 \pm 2.3$\\
$K$ (km\,s$^{-1}$) & $156.7 \pm 6.3$ & $146.7 \pm 5.9$ & $166.5 \pm 1.8$ & $170.8 \pm 1.8$ & $162.8 \pm 3.2$ & $160.8 \pm 3.1$\\
$a\,\sin{i}$ (R$_{\odot}$) & $26.3 \pm 1.0$ & $24.6 \pm 1.0$ & \multicolumn{2}{c|}{$56.8 \pm 0.4$} & $27.4 \pm 0.5$ & $27.0 \pm 0.5$ \\
$q = m_{1}/m_{2}$ & \multicolumn{2}{c|}{$0.94 \pm 0.05$} & \multicolumn{2}{c|}{$ 1.03 \pm 0.02$} & \multicolumn{2}{c}{$0.99 \pm 0.02$}\\
$\omega (^{\circ})$ & \multicolumn{2}{c|}{$257.4 \pm 14.0$} & \multicolumn{2}{c|}{$252 \pm 6$} & \multicolumn{2}{c}{$271.5 \pm 7.5$}\\
$m\,\sin^3{i}$ (M$_{\odot}$) & $11.7 \pm 1.2$ & $12.5 \pm 1.3$ & $17.1 \pm 0.6$ & $16.6 \pm 0.6$ & $14.6 \pm 0.7$ & $14.8 \pm 0.8$\\
$R_{\rm RL}/(a_1 + a_2)$ & $0.37 \pm 0.01$ & $0.38 \pm 0.01$ &  &  & $0.38 \pm 0.01$ & $0.38 \pm 0.01$\\
$R_{\rm RL}\,\sin{i}$ & $19.0 \pm 0.6$ & $19.6 \pm 0.6$ &  &  & $20.6 \pm 0.3$ & $20.7 \pm 0.3$\\
$\sigma_{\rm fit}$ (km\,s$^{-1}$) & \multicolumn{2}{c|}{3.18} & \multicolumn{2}{c|}{} & \multicolumn{2}{c}{1.86}\\
\hline 
\end{tabular}
\end{center}
\tablefoot{$T_0$ refers to the time of periastron. The values $\gamma$, $K,$ and $a\,\sin{i}$ denote the apparent systemic velocity, the semi-amplitude of the RV curve, and the projected separation between the centre of the star and the centre of mass of the binary system, respectively. The value $\omega$ is the longitude of periastron measured from the ascending node of the orbit of the primary. The value $R_{\rm RL}$ stands for the radius of a sphere with a volume equal to that of the Roche lobe computed according to the formula of \citet{Eggleton83}. All error bars indicate 1$\sigma$ uncertainties.}
\end{table*}

\subsection{HD~206267Aa}
For the inner binary system of HD~206267A, we derived a new orbital solution based on the RVs of the strongest absorption lines that are essentially free from blends with other features. For each observation, the RVs of the primary and secondary stars were computed as the mean of the RVs of the He\,{\sc i} $\lambda\lambda$\,4471, 5876, 7065 and He\,{\sc i} + {\sc ii} $\lambda$\,4026 lines determined by a multi-Gaussian fit on the observed spectra. Broader spectral features such as He\,{\sc ii} and H$\gamma$ lines of the primary, secondary and tertiary stars were too heavily blended to be measured. Again, we used the effective wavelengths of \citet{Underhill95} and \citet{Ninkov}. The results are provided in Table\,\ref{journal_206267}. The uncertainties on these RVs are about 10 -- 20\,km~s$^{-1}$ for most observations, but exceed 30\,km\,s$^{-1}$ for some spectra, indicated by the colons in Table\,\ref{journal_206267}. We also obtained a mean RV of $-7.0 \pm 7.8$\,km\,s$^{-1}$ for the tertiary component of the system. The dispersion of this value is comparable to the typical uncertainty on the individual tertiary RVs. All tertiary RV data points fall within 2\,$\sigma$ of the mean. 

A Fourier analysis of the RV(Aa1) and RV(Aa2) data yielded a highest peak around $\nu = 0.269502 \pm 0.000015$\,d$^{-1}$, i.e.\ $P_{\rm orb} = 3.710534 \pm 0.000208$\,days, which is close to the period of $P_{\rm orb} = 3.709838 \pm 0.000010$\,days proposed by \citet{Stickland95}, although the values do not agree within their error bars. We also performed a Fourier analysis on our RVs combined with those measured by \citet{Stickland95} and \citet{Burkholder97}. This time, the highest peak was found at $\nu=0.269558 \pm 0.000008$\,d$^{-1}$, i.e.\ $P_{\rm orb} = 3.709777 \pm 0.000103$\,days, which is in better agreement with the value of \citet{Stickland95}, but does not agree with the value obtained from our data only.

\begin{figure*}
\begin{center}
\begin{minipage}{8.5cm}
\resizebox{8.5cm}{!}{\includegraphics{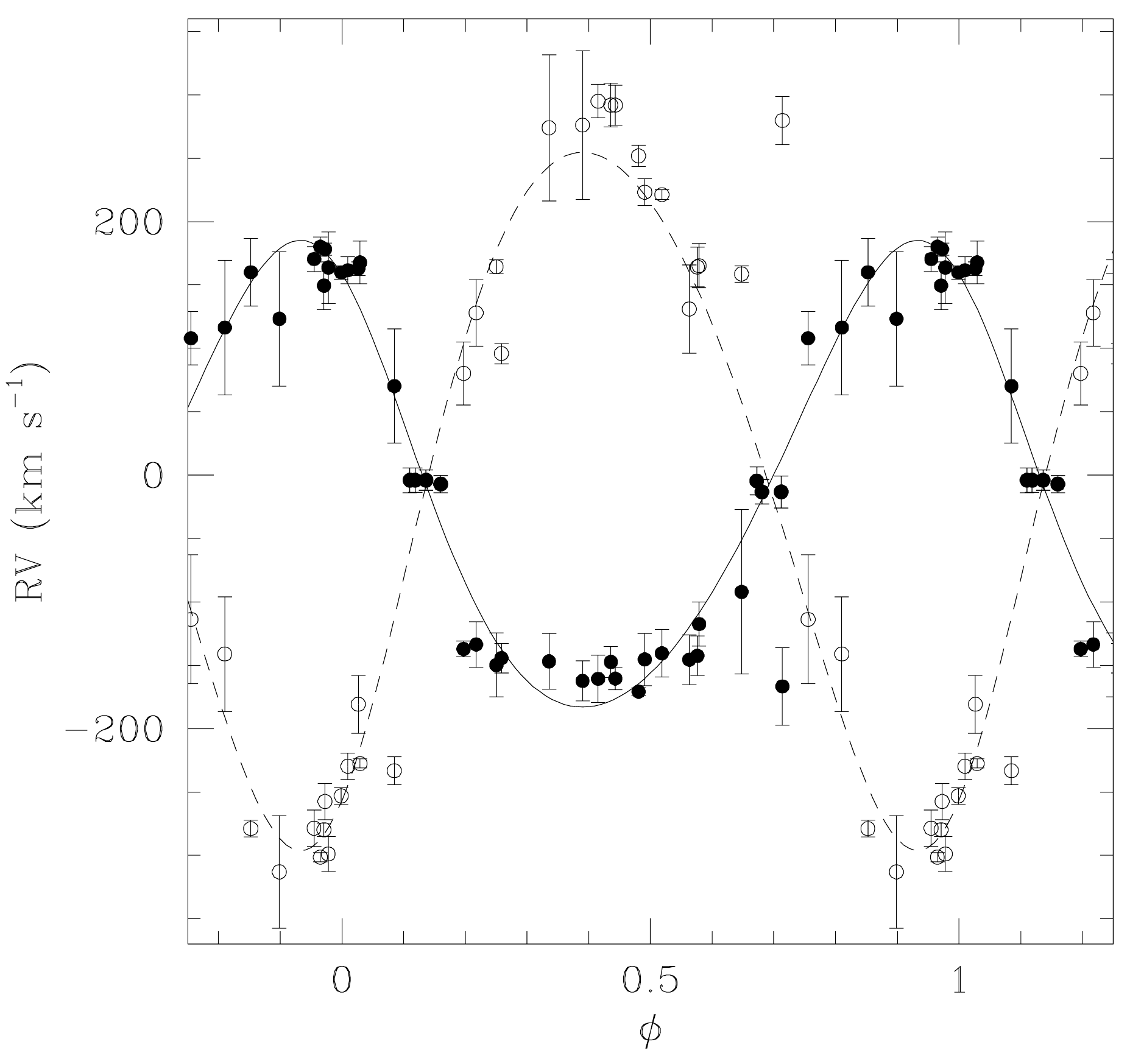}}
\end{minipage}
\hfill
\begin{minipage}{8.5cm}
\resizebox{8.5cm}{!}{\includegraphics{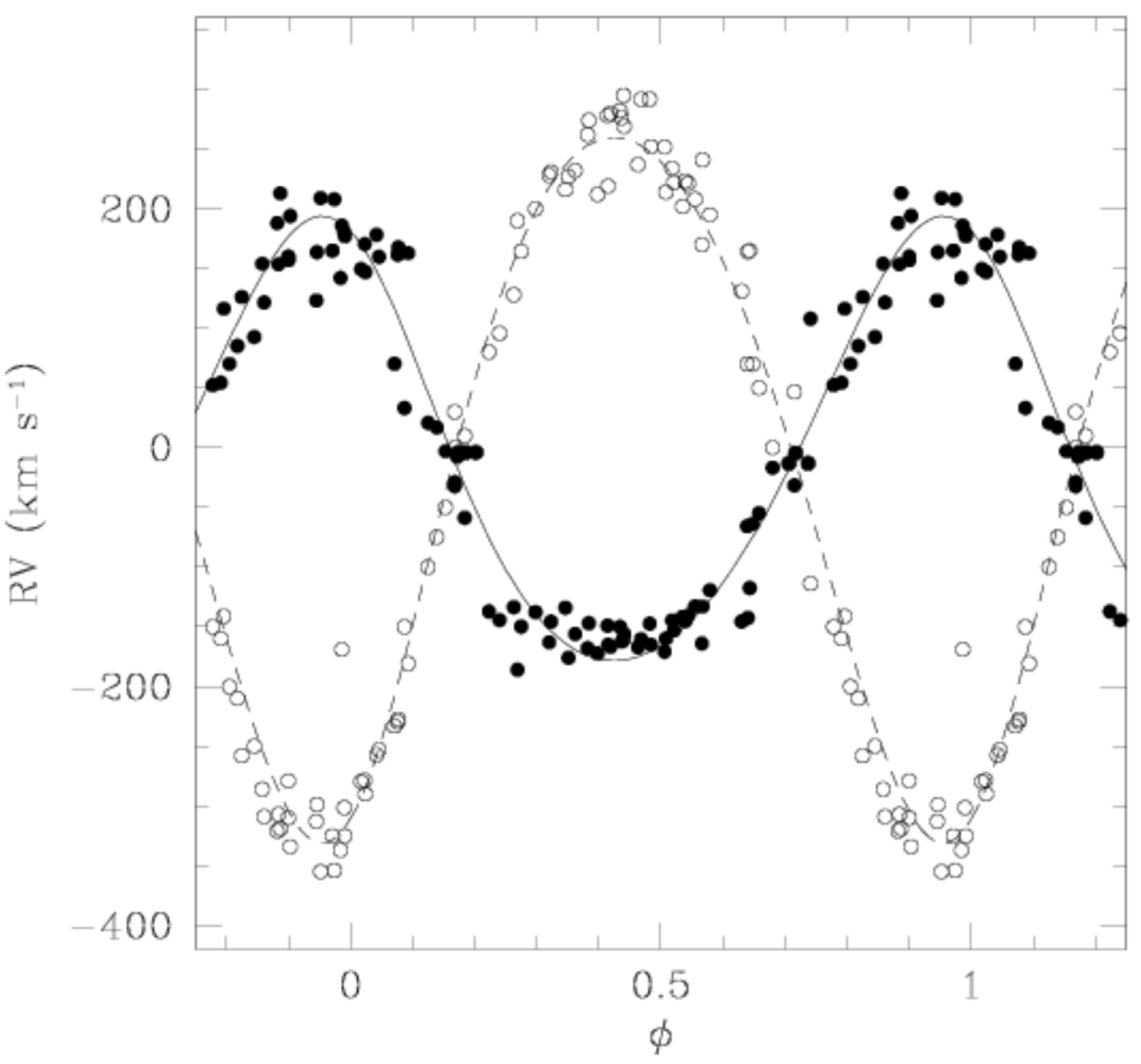}}
\end{minipage}
\end{center}
\caption{Radial velocities of HD~206267Aa1 (filled circles) and HD~206267Aa2 (open circles). The left panel shows the RVs corresponding to our new measurements (Table\,\ref{journal_206267}), assuming a period of 3.710534\,days, while the right panel shows our measurements along with those taken by \citet{Stickland95} and \citet{Burkholder97}, assuming a period of 3.709777\,days. The solid and dashed lines indicate the orbital solution from Table \ref{Orb_sol_206267}.
\label{orb_sol_fig_206267}}
\end{figure*}

Adopting orbital periods of 3.710534 and 3.709777 days, we then computed two orbital solutions with the LOSP code. The RVs were weighted according to their estimated uncertainties. The resulting RV curves are shown in Fig.\,\ref{orb_sol_fig_206267} and the corresponding orbital elements are provided in Table\,\ref{Orb_sol_206267}, together with those found in the previous studies of \citet{Stickland95} and \citet{Burkholder97}.

The RV amplitudes, minimum masses, mass ratios and eccentricities of both our single and combined solutions are in good agreement with those of previous works. We note however that in the orbital solution based only on our dataset, there is one observation (taken with the ELODIE instrument on HJD 2\,450\,710.4441) that seems to be at odds with the computed RV curves, while it is well integrated in the curves for our combined orbital solution. This is by far the oldest data point among our new data, which could hint at apsidal motion as the cause of this discrepancy. Indeed, tidal interactions in a close eccentric binary system can lead to a secular variation of the argument of periastron $\omega$ \citep[e.g.][and references therein]{Schmittomegadot}. We used the method of \citet{Rauw16} to fit the RV data accounting for the existence of the $\dot{\omega}$ term. In this way, we find the existing RV data of HD~206267Aa to be consistent with $\dot{\omega} = (1.24 \pm 0.84)^{\circ}$\,yr$^{-1}$. However, given the complexity of the spectrum of this triple system, additional data over a longer epoch are needed to establish the rate of the apsidal motion firmly.  
Meanwhile, we note that the sigma of the fit of the combined orbital solution is significantly better than the one of the orbital solution based on our RVs only. We thus chose to base our subsequent analysis of the inner binary system of HD~206267 on the orbital solution based on our RVs combined to those measured by \citet{Stickland95} and \citet{Burkholder97}.

\begin{table*}
\caption{Orbital solution of HD~206267Aa computed from our RV data assuming a period of 3.710534\,days, compared to the orbital solutions proposed by \citet{Stickland95} and \citet{Burkholder97}, and a solution obtained by combining our RVs with those measured by these authors.\label{Orb_sol_206267}}
\begin{tabular}{l|cc|cc|cc|cc}
\hline 
 & \multicolumn{2}{c|}{This study} & \multicolumn{2}{c|}{\citet{Stickland95}} & \multicolumn{2}{c|}{\citet{Burkholder97}} & \multicolumn{2}{c}{Combined solution}\\
 & Prim.\ (Aa1) & Seco.\ (Aa2) & Prim.\ (Aa1) & Seco.\ (Aa2) & Prim.\ (Aa1) & Seco.\ (Aa2) & Prim.\ (Aa1)& Seco.\ (Aa2)\\
\hline 
$P_{\rm orb}$ (days) & \multicolumn{2}{c|}{$3.710534 \pm 0.000208$} & \multicolumn{2}{c|}{$3.709838 \pm 0.000010$} & \multicolumn{2}{c|}{$3.709838 \pm 0.000010$ (f)} & \multicolumn{2}{c}{$3.709777 \pm 0.000103$}\\
$T_0$ (HJD$- 2\,450\,000$) & \multicolumn{2}{c|}{$5494.385 \pm 0.128$} & \multicolumn{2}{c|}{$9239.720 \pm 0.067$} & \multicolumn{2}{c|}{$9239.720 \pm 0.067$ (f)} & \multicolumn{2}{c}{$5494.503 \pm 0.069$}\\
$e$ & \multicolumn{2}{c|}{$0.1303 \pm 0.0231$} & \multicolumn{2}{c|}{$0.119 \pm 0.012$} & \multicolumn{2}{c|}{$0.119 \pm 0.012$ (f)} & \multicolumn{2}{c}{$0.1306 \pm 0.0149$}\\
$\gamma$ (km\,s$^{-1}$) & $-19.4 \pm 5.0$ & $9.9 \pm 6.3$ & $-24.8 \pm 1.4$ &  & $-10.7 \pm 9.4$ & $-6.8 \pm 8.3$ & $-14.9 \pm 3.1$ & $0.1 \pm 4.0$\\
$K$ (km\,s$^{-1}$) & $184.0 \pm 6.2$ & $275.4 \pm 9.2$ & $161.1 \pm 2.5$ & $288.0 \pm 11.5$ & $187.5 \pm 5.7$ & $307.6 \pm 3.8$ & $186.1 \pm 3.6$ & $295.5 \pm 5.7$\\
$a\,\sin{i}$ (R$_{\odot}$) & $13.4 \pm 0.4$ & $20.0 \pm 0.7$ & $11.7 \pm 0.18$ &  & $13.6 \pm 0.4$ & $22.4 \pm 0.3$ & $13.5 \pm 0.3$ & $21.5 \pm 0.4$\\
$q = m_1/m_2$ & \multicolumn{2}{c|}{$1.50 \pm 0.07$} & \multicolumn{2}{c|}{$1.79$} & \multicolumn{2}{c|}{$1.64$} & \multicolumn{2}{c}{$1.59 \pm 0.04$}\\
$\omega (^{\circ})$ & \multicolumn{2}{c|}{$31.2 \pm 12.7$} & \multicolumn{2}{c|}{$13.1 \pm 6.7$} & \multicolumn{2}{c|}{$13.1 \pm 6.7$ (f)} & \multicolumn{2}{c}{$21.3 \pm 6.9$}\\
$m\,\sin^3{i}$ (M$_{\odot}$) & $21.8 \pm 1.8$ & $14.5 \pm 1.1$ & $22.1 \pm 2.5$ & $12.3 \pm 0.8$ & $28.4 \pm 1.5$ & $17.3 \pm 1.5$ & $25.7 \pm 1.2$ & $16.2 \pm 0.7$\\
$R_{\rm RL}/(a_1 + a_2)$ & $0.41 \pm 0.01$ & $0.34 \pm 0.01$ &  &  &  &  & $0.42 \pm 0.01$ & $0.34 \pm 0.01$\\
$R_{\rm RL}\,\sin{i}$ & $13.8 \pm 0.4$ & $11.6 \pm 0.3$ &  &  &  &  & $14.7 \pm 0.2$ & $11.9 \pm 0.2$\\
$\sigma_{\rm fit}$ (km\,s$^{-1}$) & \multicolumn{2}{c|}{$6.08$} & \multicolumn{2}{c|}{} & \multicolumn{2}{c|}{} & \multicolumn{2}{c}{2.28}\\
\hline 
\end{tabular}
\tablefoot{The notations have the same meaning as in Table\,\ref{Orb_sol_17505_tab}. All error bars indicate 1$\sigma$ uncertainties. The ``(f)'' in the orbital solution derived by \citet{Burkholder97} stands for values adopted by these authors from the previous determination of \citet{Stickland95}.}
\end{table*}

\section{Preparatory analysis \label{preparation}}
Once the orbital solutions of the close binary systems had been determined, the natural step to follow would have been to disentangle the spectra of the three components of HD~17505A and HD~206267A, considering a non-moving tertiary object. This is a rather good approximation since the orbital period of the third component of HD~17505A around the inner binary is much longer than the time span of the observations. Moreover, a physical link between the inner binary of HD~206267A and the third component has not been established yet. 

However, when we applied our disentangling routine adapted to triple systems \citep{Mahy12}, we observed the appearance of artefacts in the resulting reconstructed primary, secondary, and tertiary spectra. Indeed, the wings of broad features in the resulting tertiary spectrum appear in emission and have a profile very similar to a mirrored profile of the wings of the corresponding lines in the primary and secondary spectra.

The appearance of such artefacts is inherent to the disentangling code itself. Indeed, this method is based on the Doppler shift of the lines of the different components of a system due to their orbital motion (see Sect.\,\ref{disent}). In the systems studied in this work, the wings of the lines of the (stationary) tertiary star are always partially blended with the neighbouring wings of the lines of the primary and secondary stars. This leads to ambiguities in the determination of the line profiles, which translate into the above described artefacts. 

We thus adopted a different approach, in which we first adjust the ternary spectrum via the combination of three synthetic spectra (see Sect.\,\ref{Adjustment}), and then subtract the tertiary spectrum from the actual observations.

To assess the relative contributions of each star to the combined spectrum of the triple systems, and therefore to normalize the synthetic spectra before combining them, we evaluated the dilution of prominent spectral lines in the observed spectra as follows:
\begin{equation}
\frac{l_i}{l_i + l_j + l_k}=\frac{l_i}{l_j\left(\frac{l_i}{l_j} + 1 + \frac{l_k}{l_j}\right)}
\end{equation}
with
\begin{equation}
\frac{l_i}{l_j}=\left(\frac{EW_i}{EW_j}\right)_{{\rm obs}}\left(\frac{EW_{STj}}{EW_{STi}}\right)_{{\rm mean}}
,\end{equation}
where $i$, $j,$ and $k$ represent either the primary, secondary, or tertiary stars, depending on the contribution we were calculating. The value $EW_i$ represents the equivalent width of the studied line of star $i$ referring to the combined continuum of the three stars and $EW_{STi}$ the same quantity measured on the synthetic CMFGEN spectrum of a typical single star of the same spectral type as star $i$. 

For HD~17505A, we considered the spectral types previously determined by \citet{Hillwig06}: i.e. O7.5\,V((f)) + O7.5\,V((f)) + O6.5\,III((f)). This way, we obtained that the primary, secondary, and tertiary contributions to the total flux of this star are 29\%, 34\%, and 37\%, respectively, which is close to the values of 30\%, 30\%, and 40\% inferred by \citet{Hillwig06}. For HD~206267A, we considered the spectral types previously determined by \cite{Burkholder97}: i.e. O6.5\,V((f)) + O9.5:\,V for the close binary system and O8\,V for the tertiary component as determined by \citet{Harvin03}. This way, we estimated that the primary, secondary, and tertiary contributions to the total flux amount to 60\%, 15\%, and 25\%, respectively. Our estimated light contribution of the third component is in good agreement with the $1.314 \pm 0.090$ magnitude difference between the binary system and the third component as found by \citet{Aldoretta15} with the Fine Guidance Sensor on the {\it HST}.  

\subsection{Adjustment of the ternary spectrum\label{Adjustment}}
We combined synthetic spectra of each component of the triple systems, shifted by the appropriate RVs for each observation, and scaled according to the brightness ratios inferred above. We then compared the resulting synthetic ternary spectrum to the observations of the system. To construct the synthetic spectrum of each component of the triple systems, we used the non-LTE model atmosphere code CMFGEN \citep{Hillier98}. This code solves the equations of radiative transfer and statistical equilibrium in the co-moving frame. The CMFGEN code is designed to work for both plane-parallel and spherical geometries and can be used to model Wolf-Rayet and O-type stars, as well as luminous blue variables and supernovae. The CMFGEN code further accounts for line blanketing and its impact on the spectral energy distribution. The hydrodynamical structure of the stellar atmosphere is directly specified as an input to the code. A $\beta$ law is used to describe the stellar wind velocity, and a super-level approach is adopted for the resolution of the equations of statistical equilibrium. The following chemical elements and their ions were included in the calculations: H, He, C, N, O, Ne, Mg, Al, Si, S, Ca, Fe, and Ni. A photospheric structure was computed from the solution of the equations of statistical equilibrium, and was then connected to the same $\beta$ wind velocity law. We assumed a microturbulent velocity in the atmosphere varying linearly with wind velocity from 10 km\,s$^{-1}$ at the photosphere to $0.1\,v_{\infty}$ at the outer boundary. The generated synthetic spectra were then combined and compared to the spectra of the triple system to constrain iteratively the fundamental properties of the three stars.

As a starting point, we assumed that the fundamental parameters of the stars and their winds have values close to those typical for stars of the same spectral type. We thus took the surface gravities, luminosities, and effective temperatures from \citet{Martins05}, mass-loss rates and $\beta$ of the wind velocity law from \citet{Muijres12}, and wind terminal velocities from \citet{Prinja90}. 
To obtain a first approximation of the line broadening, we performed a multi-Gaussian fit of some lines on the least blended observed spectra. For HD~17505A, we used the He\,{\sc i} $\lambda\lambda$\,4471, 4922, O\,{\sc iii} $\lambda$\,5592, and C\,{\sc iv} $\lambda$\,5812 lines, whereas in the case of HD~206267A, we used the He\,{\sc i} $\lambda$\,4471, He\,{\sc i} + {\sc ii} $\lambda$\,4026, and O\,{\sc iii} $\lambda$\,5592 lines.

Starting from these first approximations, we then constrained the physical properties of each star by an iterative process because each adjustment of a given parameter leads to some modifications in the value of others, and each modification in the spectrum of one of the three stars may require some modifications in the spectrum of one of the other two. For each star, the following process was used to adjust the fit of the spectra.

The first step consists in adjusting the effective temperature. This parameter is mainly determined through the relative strengths of the He\,{\sc i} $\lambda$\,4471 and He\,{\sc ii} $\lambda$\,4542 lines \citep[e.g.][]{Artemio,Martins11}. Next comes a first approximation of the surface gravities through the width of the Balmer lines, which were approximatively reconstructed with a multi-Gaussian fit on a number of observations at different phases. The next logical step would be to adjust the wind parameters. The diagnostics of the terminal velocities, $\beta$ and the mass-loss rates are the strength of H$\alpha$, the width of He\,{\sc ii} $\lambda$\,4686 and H$\alpha$, and the strengths of H$\gamma$ and H$\delta$, respectively, while the clumping filling factors and the clumping velocity factors are based on the shape of the H$\alpha$ and H$\beta$ lines. The He\,{\sc ii} $\lambda$\,4686 and the Balmer lines of the inner binary could be polluted by some emission from a wind-wind interaction zone within the inner binary. The adjustment of the models onto the observed spectra may thus lead to an underestimate of the terminal velocities, and an overestimate of the $\beta$ of the velocity law, the clumping filling factors, the clumping velocity factors, and the mass-loss rates of the primary and secondary stars. The values obtained with such a fit should thus only be considered as lower and upper limits of the real properties of the corresponding stellar winds. 

Once the fundamental stellar parameters were established, we investigated the CNO abundances through the strengths of associated lines. We performed a normalized $\chi^{2}$ analysis to determine the best fit to selected diagnostic lines, following \citet{Martins15} and \citet{Raucq16,Raucq17}. The list of suitable lines for CNO surface abundance determination was established based on the results of \citet{Martins15} and by restricting ourselves to those lines that are present in the spectra of our stars, given their spectral types.

For HD~17505Aa, we used the C\,{\sc iii} $\lambda\lambda$\,4068-70, 4156, 4163, 4187; N\,{\sc ii} $\lambda$\,4379, and N\,{\sc iii} $\lambda\lambda$\,4511, 4515, 4525; and O\,{\sc iii} $\lambda\lambda$\,5508, 5592 lines to constrain the C, N, and O abundances, respectively. For HD~206267A, the same abundances were estimated by means of the C\,{\sc iii} $\lambda\lambda$\,4068-70; N\,{\sc ii} $\lambda$\,4379, N\,{\sc iii} $\lambda\lambda$\,4511, 4515, 4525; and O\,{\sc iii} $\lambda\lambda$\,5508, 5592 lines for the primary star. For the secondary star, the C\,{\sc iii} $\lambda\lambda$ 4163, 4187 and N\,{\sc iii} $\lambda\lambda$\,4530, 4535 lines were included in addition to the line list for the primary.

\begin{figure}[h]
\begin{center}
\resizebox{9cm}{!}{\includegraphics[bb=18bp 180bp 430bp 710bp,clip]{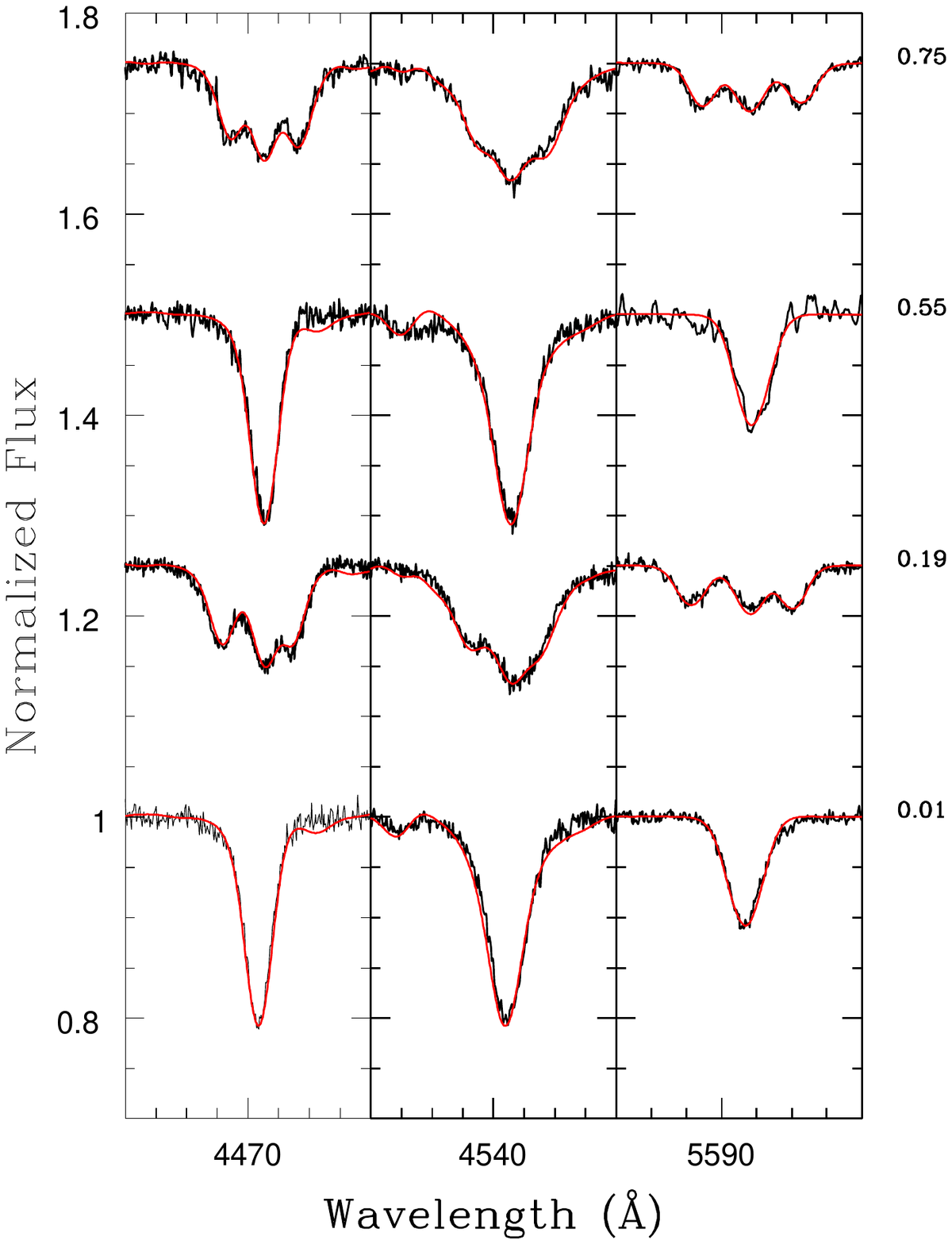}}
\end{center}
\caption{He\,{\sc i} $\lambda$ 4471 (left panel), He\,{\sc ii} $\lambda$ 4542 (middle panel), and O\,{\sc iii} $\lambda$ 5592 lines (right panel) of the recombined synthetic spectrum (red) of the triple system HD~17505A for different observations (black). The phases of the different observations, according to the ephemerides of Table\,\ref{Orb_sol_17505_tab}, are presented on the right side of the figure. The normalized spectra are shifted upwards to improve the clarity of the plot. \label{fig_recombine_17505}}
\end{figure}

\begin{figure}[h]
\begin{center}
\resizebox{9cm}{!}{\includegraphics{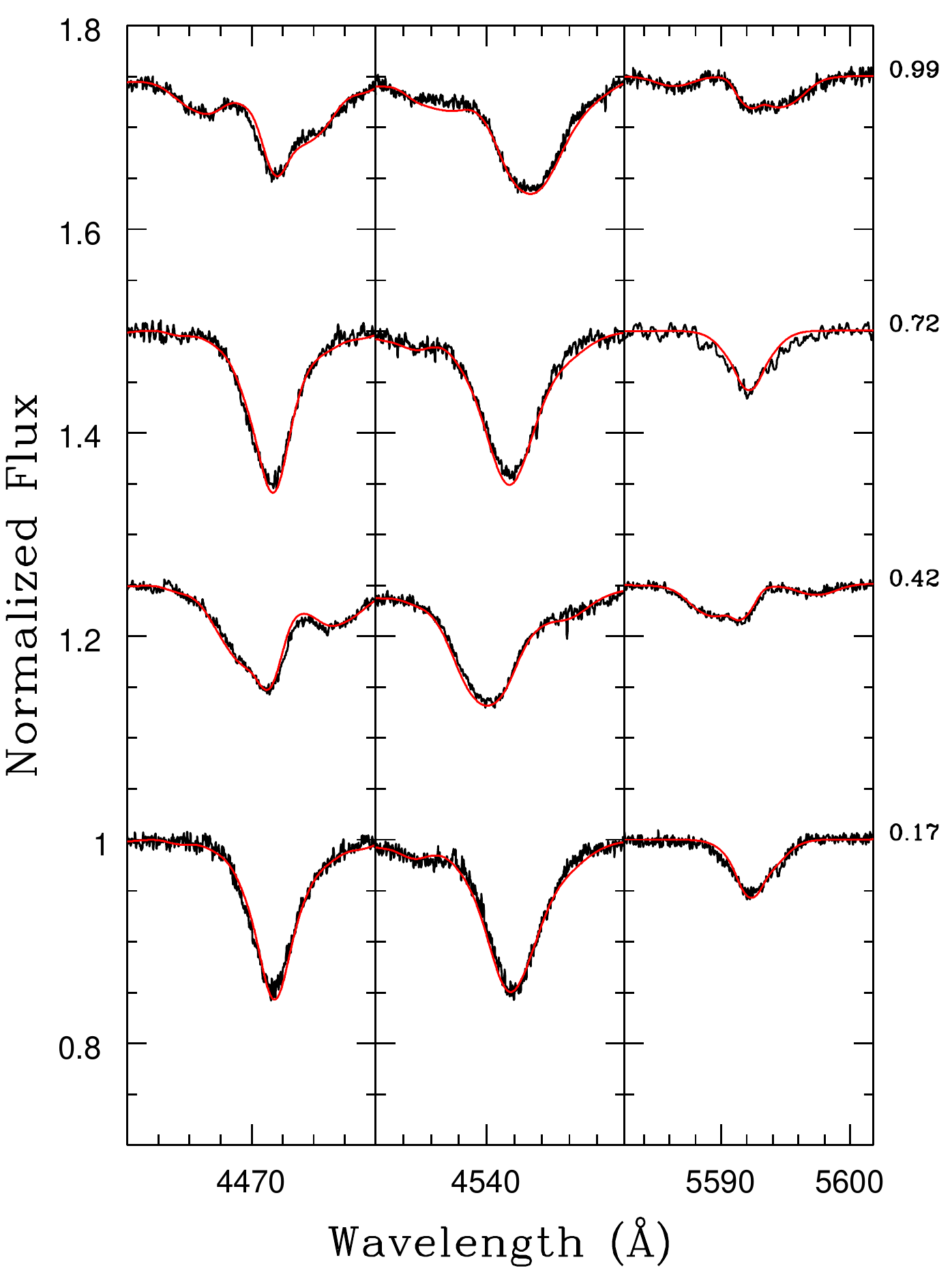}}
\end{center}
\caption{Same as Fig.\,\ref{fig_recombine_17505}, but for HD~206267A. The phases of the different observations are computed with the ephemerides of Table\ \ref{Orb_sol_206267}. \label{fig_recombine_206267}}
\end{figure}

In Figs.\,\ref{fig_recombine_17505} and \ref{fig_recombine_206267} we show parts of the observed spectra of HD~17505A and HD~206267A, respectively, at different orbital phases along with the corresponding recombined synthetic spectra of the triple systems.


\subsection{Spectral disentangling of the inner binaries \label{disent}}
Once we obtained a good fit of the triple system for all the observations, we subtracted the synthetic tertiary spectrum from each observation to recover the spectra of the inner binary. After removal of the third object, we then treated the spectra with our disentangling code \citep{Mahy12,Rauw16} based on the method introduced by \citet{GL}, in the same way as for HD~149404 \citep{Raucq16} and LSS~3074 \citep{Raucq17}. In the spectral disentangling procedure, the individual spectra are reconstructed in an iterative way by averaging the observed spectra shifted into the frame of reference of one binary component after having subtracted the current best approximation of the companion's spectrum shifted to its current estimated RV. Improved estimates of the RVs of the stars can then be evaluated by cross-correlating the residual spectra, obtained after subtracting the companion's spectrum, with a synthetic spectrum. The same steps are performed alternately for the primary and secondary star. This method allows in principle to reconstruct simultaneously
the primary and secondary spectra and determine their RVs. In the present case, the RVs of the binary components were kept fixed to the values used for deriving the orbital solutions. 

As any spectral disentangling procedure, our method also has its limitations \citep[see also][]{PH10,Mahy17}. In the present case, the most severe problem arises from the difficulties to reconstruct accurately the wings of the lines notably due to the blending with the lines of the tertiary component. This problem was at least partially solved via the subtraction of the synthetic tertiary spectrum in Sect.\,\ref{Adjustment}. Another issue stems from the fact that small residual normalization uncertainties lead to low-frequency oscillations in the reconstructed spectra. Therefore, we re-normalized the disentangled spectra using the same continuum windows as for the normalization of the original spectra. Moreover, any spectral feature that does not follow the orbital motion of either star (e.g.\ emission from a wind interaction zone located between the stars) is erroneously distributed between the primary and secondary spectra. In the systems investigated here, we did not observe any such emission. Finally, the intrinsic spectra of the stars could be variable as a function of orbital phase (e.g. because of the temperature distribution over the surface of a distorted star). However, \citet{Palate} have shown that, as long as the binary is not in a contact configuration, spectral disentangling provides a very good description of the mean spectra averaged over the individual stellar surfaces.

\subsection{Spectral types}
Once we obtained the individual reconstructed spectra of the primary and secondary components of the close binary systems, we determined their spectral types. To do so, we applied Conti's quantitative classification criteria for O-type stars \citep{Conti71,Conti77,vdHucht96} and we used a comparison with the catalogue of \citet{Walborn90}. For HD~17505Aa, we classified the primary and secondary as O7V((f)) stars, in excellent agreement with the spectral types given by \citet{Hillwig06} who proposed an O7.5\,V((f)) classification for both stars. Applying the same approach to HD~206267Aa, we found that the primary and secondary are of spectral types O5.5\,V((f)) and O9.5\,V, respectively. Our determination of the
spectral type of the secondary exactly matches that proposed by \citet{Burkholder97}, but we obtain a slightly earlier spectral type for the primary component (O5.5 versus\ O6.5 in the study of these authors). 


\subsection{Brightness ratio\label{Brightness-ratio}}
Whilst the spectral disentangling yields the strength of the lines in both primary and secondary spectra compared to the combined continuum, it does not permit us to directly establish the relative strengths of the individual continua. We thus used a similar technique as above based on the dilution of the spectral lines. We measured the equivalent widths of a number of selected spectral lines on the reconstructed spectra of the primary and secondary stars, but referring to the combined continuum of the two stars.

The results of our measurements are listed in Table\,\ref{tab_EW_17505} and Table\,\ref{tab_EW_206267}. From these numbers, we calculated the brightness ratio as
\begin{equation}
\frac{l_1}{l_2}=\left(\frac{EW_1}{EW_2}\right)_{\rm obs}\left(\frac{EW_{ST2}}{EW_{ST1}}\right)_{\rm mean}
\end{equation}
with the same notations as in the previous section. In the case of HD~17505Aa, since the spectral types of the primary and secondary are identical, the second term of the equation reduces to 1, and the brightness ratio of the two stars can be simply evaluated as the ratio of their respective equivalent widths, averaged over the selected lines. In this way, we obtained for HD~17505Aa a primary over secondary brightness ratio of $0.88 \pm 0.09$, which is very close to what we inferred from our preliminary study of the contributions of the different components to the observed spectra of the triple system, based on the multi-Gaussian fit, i.e.\ $l_1/l_2 = 0.85$. Our estimate is slightly lower than the brightness ratio of 1.00 proposed by \citet{Hillwig06}. For HD~206267Aa, we found a mean primary over secondary brightness ratio of $4.75 \pm 1.79$, coherent, within the error bars, with what we inferred from our preliminary study based on the multi-Gaussian fit, i.e.\ $l_1/l_2 = 4.00$.

\begin{table}
\caption{Determination of the brightness ratio of the close binary system HD~17505Aa from the dilution of prominent lines.\label{tab_EW_17505}}
\begin{center}
\begin{tabular}{c|cc|c}
\hline 
Line & \multicolumn{2}{c|}{EW (\AA)} & $l_1/l_2$\\
 & Primary (Aa1) & Secondary (Aa2) & \\
\hline 
He\,{\sc i} + {\sc ii} $\lambda$\,4026 & 0.11  & 0.15 & 0.73\\
He\,{\sc ii} $\lambda$\,4200 & 0.09  & 0.10 & 0.94\\
H$\gamma$ & 0.51 & 0.52 & 0.98\\
He\,{\sc i} $\lambda$\,4471 & 0.12 & 0.14 & 0.88\\
He\,{\sc ii} $\lambda$\,4542 & 0.14 & 0.17 & 0.87\\
\hline 
\multicolumn{3}{r|}{Mean value} & $0.88 \pm 0.09$ \\
\hline
\end{tabular}
\end{center}
\end{table}

\begin{table}
\caption{Determination of the brightness ratio of the close binary system HD~206267Aa from the dilution of prominent lines.\label{tab_EW_206267}}
\begin{center}
\tiny
\begin{tabular}{l|cccc|c}
\hline 
Line & \multicolumn{4}{c|}{EW (\AA)} & \multicolumn{1}{c}{$l_1/l_2$}\\
 & \multicolumn{2}{c}{Observations} & \multicolumn{2}{c|}{Synth.\ spectra} & \\
 & Pri.\ (Aa1) & Sec.\ (Aa2) & O5.5 & O9.5 & \\
\hline 
He\,{\sc i} + {\sc ii} $\lambda$\,4026 & 0.27 & 0.09 & 0.74 & 1.02 & 4.22\\
He\,{\sc ii} $\lambda$\,4200 & 0.32 & 0.02 & 0.76 & 0.16 & 4.18\\
H$\gamma$  & 1.27 & 0.19 & 2.11 & 2.52 & 8.16 \\
He\,{\sc i} $\lambda$\,4471 & 0.18 & 0.11 & 0.41 & 1.04 & 4.29\\
He\,{\sc ii} $\lambda$\,4542 & 0.48 & 0.02 & 0.92 & 0.13 & 2.87\\
\hline
\multicolumn{5}{r|}{Mean value} & $4.75 \pm 1.79$ \\
\hline 
\end{tabular}
\end{center}
\end{table}

The compilation of \citet{Reed} yields a mean $V$ magnitude of $7.07 \pm 0.02$ and a $B-V$ colour index of $0.40$ for the entire HD~17505 system. \citet{MaizApellaniz} measured an angular separation of $2.153 \pm 0.016$\,arcsec and a magnitude difference of 1.70 between the A and B components. For the HD~17505A triple system, we thus obtained $V = 7.27 \pm 0.02$. Because the intrinsic $\left(B-V\right)_0$ of O7 stars is $-0.27$ \citep{Martins06}, we inferred an extinction $A_V$ of $2.07 \pm 0.01$, assuming $R_V = 3.1$. For a distance of $\sim 1.9$\,kpc \citep{Hillwig06}, we derived an absolute magnitude $M_V = -6.19 \pm 0.11$ for the triple system, where we included a 10\% error on the distance. Accounting for a flux contribution of 37\% by the tertiary component and using our above-determined $l_1/l_2$ brightness ratio, we then calculated individual absolute magnitudes of $M_V^{P} = -4.87 \pm 0.13$ and $M_V^{S} = -5.00 \pm 0.12$.

The mean $V$ magnitude of HD~206267A as evaluated from the measurements compiled  by \citet{Reed} is $5.67 \pm 0.06$, and the mean ($B-V$) colour is $0.21 \pm 0.01$. The large dispersion of the $V$ magnitudes could hint at photometric variability\footnote{There are six magnitude determinations of this star in the compilation of \citet{Reed}: three at $V = 5.62$, one at 5.70, one at 5.71 and another one at 5.74.}. {\it Hipparcos} photometric measurements indeed confirm the presence of variability. A Fourier analysis using the \citet{HMM} method leads to the highest peak at $\nu = 0.3398$\,d$^{-1}$ with an amplitude of 0.013\,mag. This corresponds to a period of 2.94\,days, which is not compatible with the orbital period of HD~206267Aa, but is close to our best estimate of the rotation period of the secondary. We thus conclude that, whilst there is probably low-level photometric variability, it is not due to eclipses in the inner binary system. Adopting a mean ($B-V$)$_0$ of $-0.27$ for the system \citep{Martins06}, we determined an extinction $A_V$ of $1.49 \pm 0.03$, assuming $R_V = 3.1$. With a distance modulus of $9.9 \pm 0.5$ \citep{Burkholder97}, we inferred an absolute magnitude $M_ V = -5.73 \pm 0.50$ for the system. Using the previously determined brightness ratio, we then calculated individual absolute magnitudes of $M_V^{P} = -5.21 \pm 0.51$ and $M_V^{S} = -3.52 \pm 0.61$. Based on a correlation between the strength of interstellar Ca\,{\sc ii} lines and distance, \citet{Megier} estimated a slightly larger distance modulus of $10.11 \pm 0.38$. This estimate leads to $M_V^{P} = -5.42 \pm 0.39$ and $M_V^{S} = -3.73 \pm 0.51$

The reconstructed normalized primary and secondary optical spectra of both systems are shown and discussed in Sect.\,\ref{results}.

\section{Spectral analysis \label{analysis}}
\subsection{Rotational velocities and macroturbulence}
The reconstructed individual spectra also allowed us to estimate the values of the projected rotational velocities and macroturbulence of the primary and secondary stars. To assess the projected rotational velocity, we applied a Fourier transform method \citep{SimonDiaz07,Gray08} to the profiles of isolated lines in the disentangled spectra. Since our spectra reveal very few suitable metallic lines (basically O\,{\sc iii} $\lambda$\,5592 is the only one, and this line is not available in the secondary of HD~206267Aa), we also considered the He\,{\sc i} $\lambda\lambda$\,4922, 5015 lines. We note that these two lines are less affected by Stark broadening than other He\,{\sc i} lines present in the optical spectrum of O-type stars. The results are summarized in Table\ \ref{tab_vsini}. 
\begin{table}
\caption{Projected rotational velocities ($v\,\sin{i}$ in km\,s$^{-1}$) of the primary and secondary components of HD~17505Aa and HD~206267Aa.\label{tab_vsini}}
\begin{tabular}{c|cc|cc}
\hline 
 & \multicolumn{2}{c|}{HD~17505Aa} & \multicolumn{2}{c}{HD~206267Aa}\\
Line & Prim. & Sec. & Prim. & Sec.\\
\hline 
He\,{\sc i} $\lambda$\,4922 & 49 & 55 & 166 & 99\\
He\,{\sc i} $\lambda$\,5016 & 53 & 57 & 195 & 96\\
O\,{\sc iii} $\lambda$\,5592 & 62 & 62 & 171 & -\\
\hline 
Mean value & $54.7 \pm 5.4$ & $58.0 \pm 2.9$ & $177.3 \pm 12.7$ & $97.5 \pm 1.5$\\
\hline 
\end{tabular}
\end{table}

To estimate the importance of macroturbulent broadening, we used the auxiliary program MACTURB of the stellar spectral synthesis program SPECTRUM v2.76 developed by \citet{Gray10} and based on the radial-tangential anisotropic macroturbulent broadening formulation of \cite{Gray08}. The MACTURB program provided us the following values: 60 and 65\,km\,s$^{-1}$ for the primary and secondary stars of HD~17505Aa, respectively, as well as 80 and 120\,km\,s$^{-1}$ for the primary and secondary stars of HD~206267Aa, respectively. However, since most of our measurements are based on He\,{\sc i} lines, these numbers should only be considered as upper limits on the actual value of the macroturbulence. 

Both rotational and additional broadening effects were applied on the synthetic CMFGEN spectra before comparing the latter with the reconstructed spectra of the primary and secondary stars of both systems.

\subsection{Fit of the separated spectra with the CMFGEN code\label{results}}
In this section, we undertake a quantitative spectral analysis of the reconstructed spectra by means of the CMFGEN model atmosphere code \citep{Hillier98}. In addition to the formal fitting errors, our parameters could be affected by possible systematic errors. The systematic errors of model atmosphere codes are difficult to estimate in an absolute way, but some insight comes from comparison of different model atmosphere codes applied to the same stars. 
\citet{Massey13} compared the best-fit parameters for a set of 10 Magellanic Cloud O-stars obtained with the CMFGEN and FASTWIND \citep{FASTWIND,Puls} model atmosphere codes. These authors reported differences in $T_{\rm eff}$ of about 1000\,K that are larger than the uncertainties in determining these parameters with each code, but they found no systematic difference. On the other hand, they found a systematic difference of 0.12\,dex between the $\log{g}$ values obtained with these codes. More recently, \citet{Holgado} compared the results of their FASTWIND analyses against the CMFGEN analyses available in the literature. For about 30 Galactic O-type stars in common, they found that the CMFGEN parameters yield $T_{\rm eff}$ that are lower by $\sim 800$\,K and $\log{g}$ higher by about 0.09\,dex. These results contrast with the analysis of 14 Galactic O-type stars by \citet{Berlanas}. These authours found no systematic difference in $T_{\rm eff}$ or $\log{g}$ between their CMFGEN and FASTWIND analyses. In their study of rapidly rotating Galactic O-type stars, \citet{Cazorla} compared the parameters inferred with the CMFGEN and DETAIL/SURFACE \citep{BG} codes for three O9 -- O9.5 stars. Differences on $T_{\rm eff}$ and $\log{g}$ were found to be less than 500\,K and 0.1\,dex, respectively. Moreover, \citet{Cazorla} found no significant difference between the He and CNO abundances determined with both codes.

The results obtained may also depend on the wavelength domain considered for the spectral analysis. For instance, \citet{Berlanas} emphasized the importance of the H$\gamma$ line to correctly infer the $\log{g}$ of O-type stars. Whilst this line is included in our analysis, we stress that its wings could be subject to uncertainties due to the spectral disentangling. To circumvent this problem, we have thus adopted a somewhat different approach (see Sects.\,\ref{Results-HD-17505} and \ref{Results-HD-206267}).
\subsubsection{HD~17505Aa\label{Results-HD-17505}}
Using the reconstructed primary and secondary spectra of HD~17505Aa, the brightness ratio, rotational velocities, and macroturbulence velocities, we improved our determination of the fundamental properties of both stars with the CMFGEN code. We followed the same procedure as described in Sect.\,\ref{Adjustment} to constrain the different stellar parameters. 

Unfortunately, the Balmer lines are too broad to be correctly recovered by the disentangling process, and their wings are thus not reliable to adjust the surface gravities. We instead used an iterative process to constrain the luminosities together with these gravities. From the first estimate of $\log{g}$ given by the fit of the triple system, performed in Sect.\,\ref{Adjustment}, and our determination of the effective temperatures, we inferred the bolometric corrections \citep{Lanz03} and hence the individual bolometric luminosities, using the absolute $V$ magnitudes of the components derived in Sect.\,\ref{Brightness-ratio}. These bolometric luminosities and the effective temperatures then permitted us to compute the ratio of the stellar radii $\frac{R_{P}}{R_{S}}$. Together with the assumed surface gravities, this ratio yielded the spectroscopic mass ratio $\frac{M_{P}}{M_{S}}$, which we compared to the dynamical mass ratio inferred from the orbital solution (Sect.\,\ref{New-orbital-solutions}). The difference between these mass ratios then resulted in a revised estimate of the surface gravities. This iterative process was repeated until the spectroscopic and dynamical mass ratios agreed with each other, and the CMFGEN synthetic spectra produced for the new surface gravities matched the observations as well as possible.

\begin{figure*}
\begin{center}
\resizebox{15cm}{!}{\includegraphics{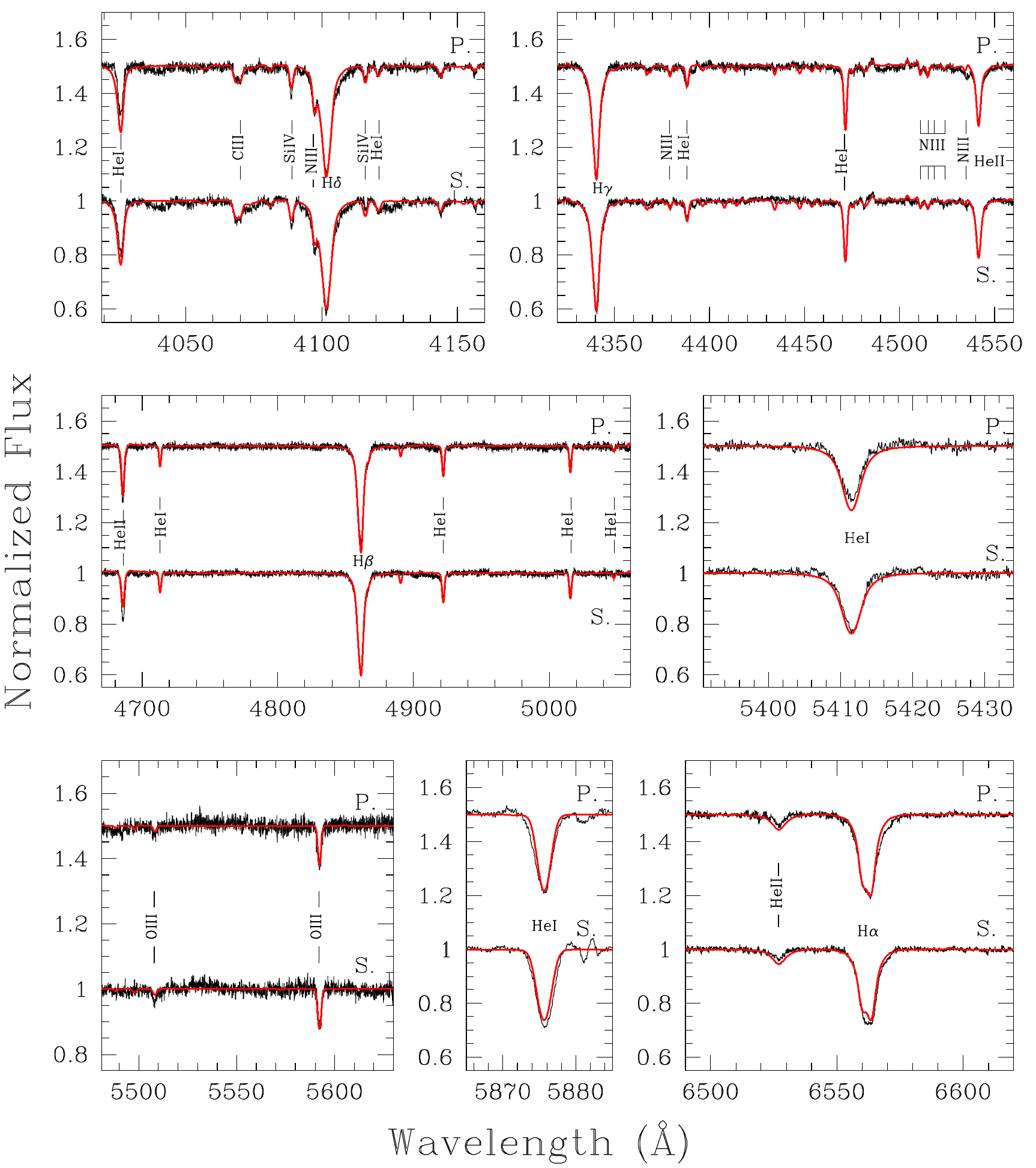}}
\end{center}
\caption{Part of the normalized disentangled spectra of the primary (P, shifted upwards by 0.5 continuum units) and secondary (S) stars of HD~17505Aa along with the best-fit CMFGEN model spectra (red). \label{fitCMFGEN_17505}}
\end{figure*}

Figure\,\ref{fitCMFGEN_17505} shows the best fit of the optical spectra of the primary and secondary stars of HD~17505Aa obtained with CMFGEN, and Table\,\ref{tableCMFGEN_17505} lists the corresponding stellar parameters. Table\,\ref{abondancesCMFGEN_17505} compares the chemical abundances of these best-fit models with the solar abundances taken from \citet{Asplund09}.

Figure \ref{fitCMFGEN_17505} shows that the H, He, C, and N lines are well reproduced by the models for both stars. The case of oxygen is  more problematic. Indeed, while the O\,{\sc iii} $\lambda\lambda$\,5508, 5592 lines are well adjusted, several other lines (e.g.\ O\,{\sc iii} $\lambda\lambda$\,4448, 4454, 4458) are present in the synthetic CMFGEN spectra of both stars, but are neither visible in the separated spectra, nor in the observed spectra before the disentangling procedure. We encountered similar problems with the same lines in the case of HD~149404 \citep[see][]{Raucq16}.

\begin{table}
\caption{Best-fit CMFGEN model parameters of the primary and secondary stars
of HD~17505Aa.}
\begin{center}
\begin{tabular}{c|cc}
\hline 
 & Prim. & Sec.\\
\hline 
$R$ (R$_{\odot}$)  & $9.7 \pm 0.8$ & $10.4 \pm 0.9$ \\
$M$ (M$_{\odot}$)  & $19.4 \pm 7.4$ & $21.8 \pm 8.3$ \\
$T_{\rm eff}$ (kK)  & $37.0 \pm 1.5$ & $36.7 \pm 1.5$ \\
$\log{\frac{L}{L_{\odot}}}$  & $5.20 \pm 0.05$ & $5.25 \pm 0.05$ \\
$\log{g}$ (cgs)  & $3.75 \pm 0.15$ & $3.74 \pm 0.15$ \\
BC & $-3.40$ & $-3.38$\\
\hline
$\beta$  & $\leq 1.07$ & $\leq 1.07$ \\
$v_{\infty}$ (km\,s$^{-1}$)  & $\geq 2200$ & $\geq 2500$\\
$\dot{M}$ (M$_{\odot}$\,yr$^{-1})$  & $\leq 7.3 \times 10^{-8}$ & $\leq 1.65 \times 10^{-7}$\\
$f_{1}$ & $\leq 0.1$ & $\leq 0.2$\\
$f_{2}$ (km\,s$^{-1}$)  & $\leq 250$ & $\leq 180$\\
\hline 
\end{tabular}
\end{center}
\tablefoot{The quoted errors correspond to $1\sigma$ uncertainties. The bolometric corrections are taken from \cite{Lanz03} based on our best-fit $T_{\rm eff}$ and $\log{g}$. The wind parameters, in the lower half of the table, are subject to very large uncertainties and should thus be taken with caution. The values $f_{1}$ and $f_{2}$ represent the clumping filling factor and the clumping velocity factor.\label{tableCMFGEN_17505}}
\end{table}

\begin{table}
\caption{Chemical surface abundances of the primary and secondary stars of
HD~17505Aa.}
\begin{center}
\begin{tabular}{c ccc}
\hline 
\multicolumn{1}{c }{} & Primary  & Secondary  & Sun\\
\hline 
He/H  & 0.1 (fixed)  & 0.1 (fixed) & 0.089\\
\vspace*{-2.5mm}\\
C/H  & $1.91_{-0.40}^{+0.37} \times 10^{-4}$  & $1.97_{-0.42}^{+0.39} \times 10^{-4}$  & $2.69 \times 10^{-4}$\\
\vspace*{-2.5mm}\\
N/H  & $1.37_{-0.21}^{+0.25} \times 10^{-4}$  & $9.70_{-0.84}^{+1.1} \times 10^{-5}$  & $6.76 \times 10^{-5}$\\
\vspace*{-2.5mm}\\
O/H  & $3.87_{-0.92}^{+1.2} \times 10^{-4}$  & $4.73_{-1.5}^{+2.2} \times 10^{-4}$  & $4.90 \times 10^{-4}$\\
\vspace*{-2.5mm}\\
\hline 
\end{tabular}
\end{center}
\tablefoot{Abundances are given by number as obtained with CMFGEN. The solar abundances \citep{Asplund09} are quoted in the last column. The 1$\sigma$ uncertainty corresponds to abundances that yield a normalized $\chi^{2}$ of 2.0 \citep{Martins15}. The values of the He abundances have been fixed in the models.\label{abondancesCMFGEN_17505}}
\end{table}

Combining the spectroscopic masses obtained with CMFGEN (Table \ref{tableCMFGEN_17505}) with the minimum masses obtained from the orbital solution (Table\,\ref{Orb_sol_17505_tab}), we compute an inclination of the orbit of $i \sim 57^{\circ}$. With this inclination, we can see from our determination of the stellar radii with CMFGEN compared to the radii of the Roche lobes obtained in the orbital solution (Table\,\ref{Orb_sol_17505_tab}) that both stars are well inside their Roche lobes: they have Roche lobe filling factors of $0.42 \pm 0.07$ and $0.45 \pm 0.07$, respectively. 

There is a slight enhancement of the N/O and N/C surface abundance ratios of both stars. These modest abundance modifications are fully compatible with predictions from single star evolutionary models accounting for rotational mixing, and there is no need to assume a past mass-exchange episode in the system (see also Sect.\,\ref{Evolutionary-status}). 

From the inclination we estimated, we can derive the absolute rotational velocities of 65.2 and 69.2\,km\,s$^{-1}$ for the primary and secondary stars, respectively. Combining these rotational velocities with the stellar radii obtained with CMFGEN (Table \ref{tableCMFGEN_17505}), we find that the primary and secondary stars of HD~17505Aa are in nearly perfect synchronous rotation with each other: $P_P = 7.53$, $P_S = 7.61$\,days. These estimated rotation periods are shorter than the orbital period of $P_{\rm orb} = 8.569$\,days. This could indicate that the system is not fully synchronized yet, although we stress that this could also stem from the uncertainties on the projected rotation velocities or hint at a possible misalignment between the orbital and rotational axes. If we calculate the critical rotational velocities of the stars, $v_{\rm crit} = \left(\frac{2GM}{3R_P}\right)^{1/2}$, based on our results from the CMFGEN fit, we find that the current rotational velocities of the primary and secondary stars correspond each to $0.13\times v_{\rm crit}$, which is close to the median value of birth rotational velocities for stars of similar masses, according to the study of \citet{Wolff06}. 

\subsubsection{HD~206267Aa\label{Results-HD-206267}}
\begin{figure*}
\begin{center}
\resizebox{15cm}{!}{\includegraphics[bb=60bp 150bp 550bp 710bp,clip]{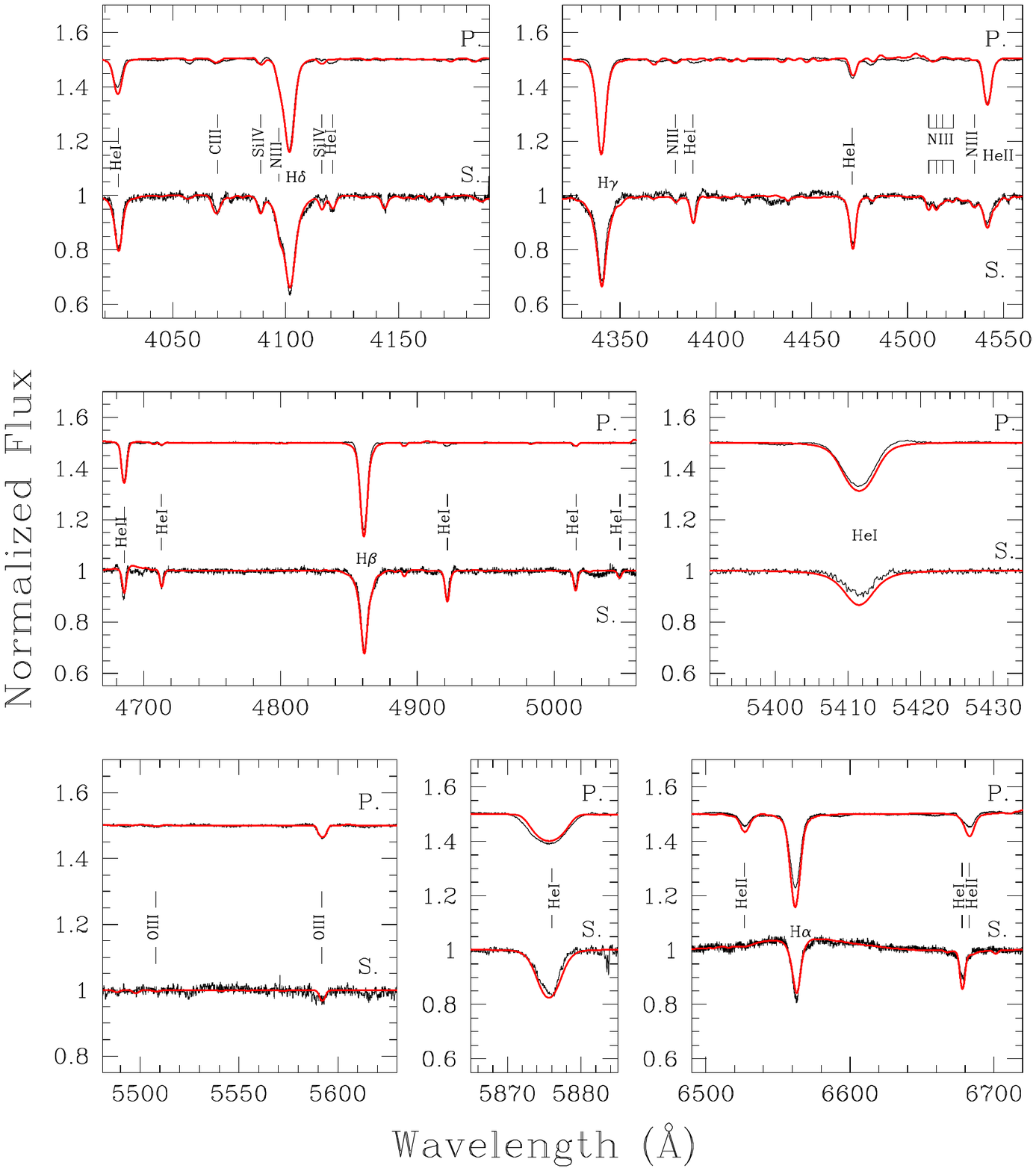}}
\end{center}
\caption{Part of the normalized disentangled spectra of the primary (P, shifted upwards by 0.5 continuum units) and secondary (S) stars of HD~206267Aa, along with the best-fit CMFGEN model spectra (red). \label{fitCMFGEN_206267}}
\end{figure*}

We applied the same method to HD~206267Aa. Here, we encountered a problem because the spectroscopic masses depend on the stellar radii which in turn are set by the stellar luminosities and thus the distance of the binary. Indeed, adopting a distance modulus of 9.9, we obtained best-fit spectroscopic masses that were lower than the minimum dynamical masses inferred from the orbital solution. We thus favour the solution corresponding to $DM = 10.11$, which leads to a better agreement between the spectroscopic masses and the minimum dynamical masses. Figure\,\ref{fitCMFGEN_206267} indicates the best fit of the optical spectra of the primary and secondary stars of HD~206267Aa obtained with CMFGEN. The H, He, C, N, and O lines are well reproduced by the models for both stars. The corresponding stellar parameters and abundances are listed in Table\,\ref{tableCMFGEN_206267} and Table\,\ref{abondancesCMFGEN_206267}, respectively.

\begin{table}[h]
\caption{Best-fit CMFGEN model parameters of the primary and secondary stars
of HD~206267Aa assuming $DM = 10.11$ \label{tableCMFGEN_206267}}
\begin{center}
\begin{tabular}{c|cc}
\hline 
 & Prim. & Sec.\\
\hline 
$R$ (R$_{\odot}$)  & $11.7 \pm 2.3$ & $5.9 \pm 1.5$\\
$M$ (M$_{\odot}$)  & $27.8 \pm 10.8$ & $17.7 \pm 8.9$\\
$T_{\rm eff}$ (kK)  & $41.0 \pm 1.5$ & $35.5 \pm 1.5$\\
$\log{\frac{L}{L_{\odot}}}$ & $5.54 \pm 0.16$ & $4.70 \pm 0.21$\\
$\log{g}$ (cgs)  & $3.75 \pm 0.15$ & $4.14 \pm 0.15$\\
BC & $-3.68$ & $-3.28$\\
\hline
$\beta$  & $\leq 0.85$ & $\leq 0.50$\\
$v_{\infty}$ (km\,s$^{-1}$)  & $\geq 2300$ & $\geq 3500$\\
$\dot{M}$ (M$_{\odot}$\,yr$^{-1})$  & $\leq 4.0 \times 10^{-8}$ & $\leq 5.02 \times 10^{-7}$\\
$f_{1}$ & $\leq 0.2$ & $\leq 0.3$\\
$f_{2}$ (km\,s$^{-1}$)  & $\leq 100$ & $\leq 300$\\
\hline 
\end{tabular}
\end{center}
\tablefoot{The notations are the same as in Table\,\ref{tableCMFGEN_17505}.}
\end{table}

\begin{table}
\caption{Chemical surface abundances of the primary and secondary stars of
HD~206267Aa.}
\begin{center}
\begin{tabular}{c ccc}
\hline 
\multicolumn{1}{c }{} & Primary  & Secondary  & Sun\\ 
\hline 
He/H  & 0.1 (fixed)  & 0.1 (fixed)  & 0.089\\
\vspace*{-2.5mm}\\
C/H  & $1.21_{-0.06}^{+0.06} \times 10^{-4}$  & $1.53_{-0.18}^{+0.17} \times 10^{-4}$ & $2.69 \times 10^{-4}$\\
\vspace*{-2.5mm}\\
N/H  & $4.15_{-0.30}^{+0.32} \times 10^{-4}$  & $2.32_{-0.82}^{+0.75} \times 10^{-4}$ & $6.76 \times 10^{-5}$\\
\vspace*{-2.5mm}\\
O/H  & $4.52_{-0.57}^{+0.80} \times 10^{-4}$  & $2.00_{-0.49}^{+0.64} \times 10^{-4}$ & $4.90 \times 10^{-4}$\\
\vspace*{-2.5mm} & \\
\hline 
\end{tabular}
\end{center}
\tablefoot{The notations are the same as in Table\,\ref{tableCMFGEN_17505}.\label{abondancesCMFGEN_206267}}
\end{table}

Comparing the spectroscopic and minimum dynamical masses (Tables\,\ref{tableCMFGEN_206267} and \ref{Orb_sol_206267}), we then estimated an orbital inclination of $i\sim76^{\circ}$. From the stellar radii determined with CMFGEN and the radii of the Roche lobes obtained in the orbital solution, we estimate mean Roche lobe filling factors of $0.77\pm0.21$ and $0.49\pm0.16$, respectively, for the primary and secondary stars of HD~206267Aa. At periastron, these Roche lobe filling factors become $0.89\pm0.22$ and $0.56\pm0.17$ for the primary and secondary stars, respectively. Whilst the secondary star is well inside its Roche lobe, the primary star, within the rather large error bars, might either be well inside its periastron Roche lobe or fill it up. 

The chemical abundances indicate an enhancement of the N/O and N/C abundance ratios at the surface of both components of the binary. We come back to these abundances in Sect.\,\ref{Evolutionary-status}.

From the inclination determined above, we derived absolute rotational
velocities of 182.7 and 100.5\,km\,s$^{-1}$ for the primary and secondary stars, respectively. By combining these rotational velocities with the stellar radii obtained with CMFGEN (Table\,\ref{tableCMFGEN_206267}), we estimated that the rotational periods of the primary and secondary stars of HD~206267Aa are 3.24 and 2.97\,days. Whilst the rotation of the primary is likely in a pseudo-synchronization state with the orbital motion \citep{Hut}, the secondary star appears to rotate at a slightly higher rate. The primary and secondary stars rotate at $0.33\times v_{{\rm crit}}$ and $0.16\times v_{{\rm crit}}$ respectively, which is close to the median value of birth rotational velocities for stars of similar masses \citep{Wolff06}.

\section{Discussion and conclusions \label{summary}}
\subsection{Evolutionary status\label{Evolutionary-status}}
In this section we compare the results of our study with single star evolutionary models from \citet{Ekstrom12} and \citet{Brott}\footnote{Whilst the models of \citet{Ekstrom12} assume Z$_{\odot} = 0.014$, those of \citet{Brott} instead assume Z$_{\odot} = 0.008$.}. We stress that the purpose of this comparison is mainly qualitative. Indeed, the evolution of the rotational angular momentum in these single star theoretical models is most probably not representative of the stars in our study. In the close binaries studied here, tidal coupling between the angular momentum of the orbital motion and that of the stellar spins affects the evolution of rotation rates. Therefore, it would be a mere coincidence if one of our stars were entirely described by these models. Nevertheless, this comparison can bring us some interesting pieces of information about the properties of our targets. 
\subsubsection{HD~17505Aa}
\begin{figure*}
\begin{minipage}{6cm}
\resizebox{6cm}{!}{\includegraphics{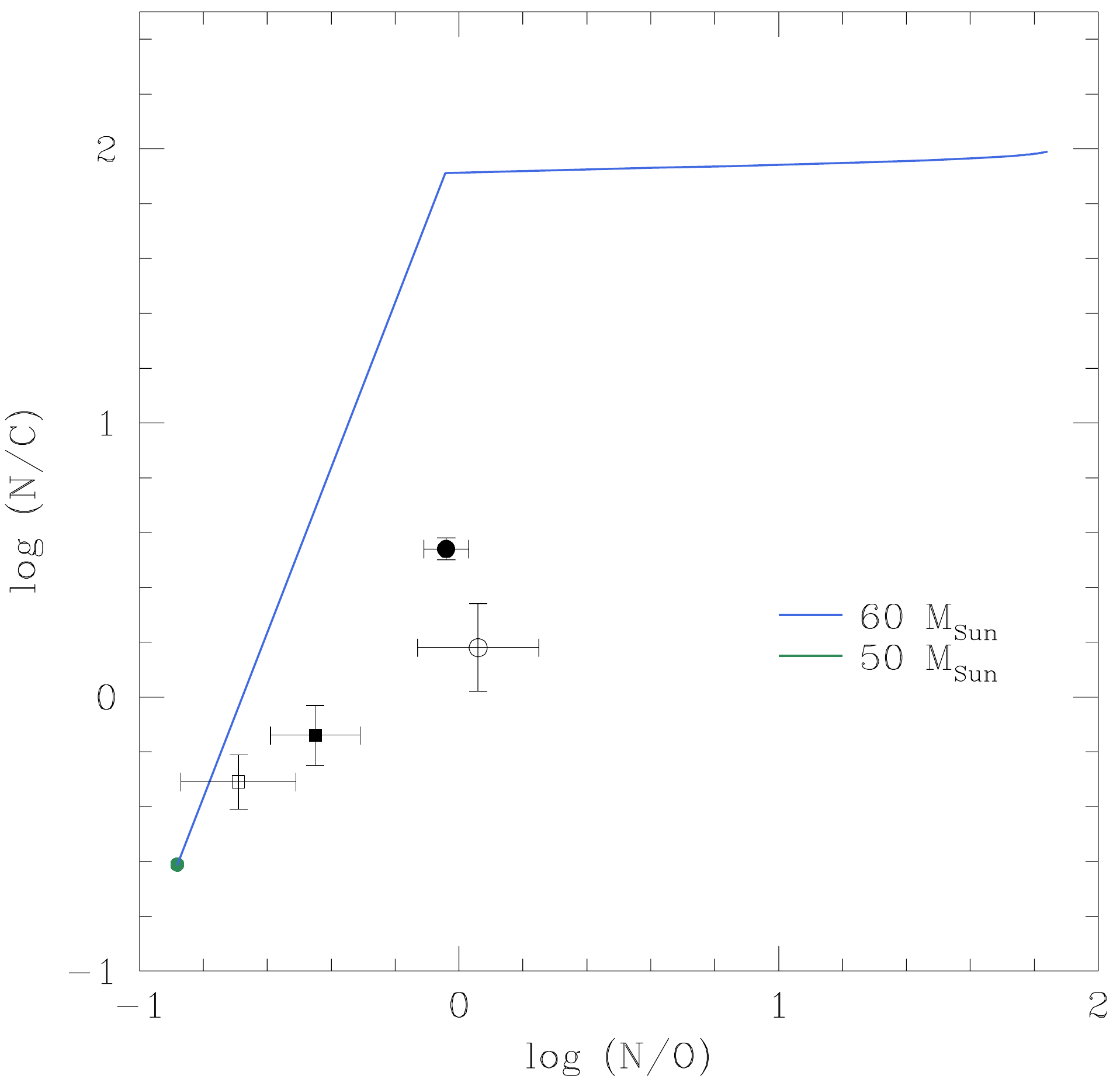}}
\end{minipage}
\hfill
\begin{minipage}{6cm}
\resizebox{6cm}{!}{\includegraphics{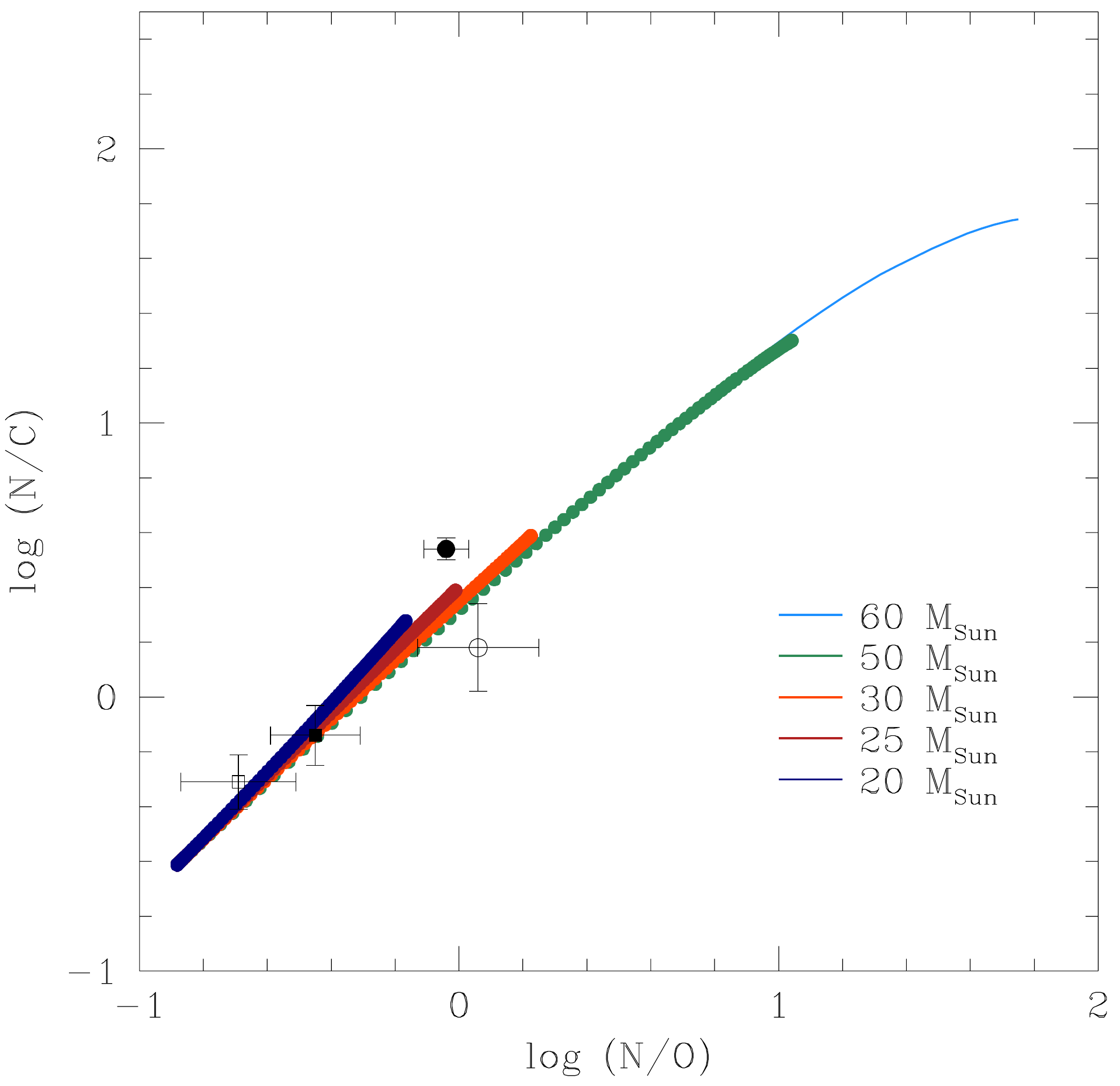}}
\end{minipage}
\hfill
\begin{minipage}{6cm}
\resizebox{6cm}{!}{\includegraphics{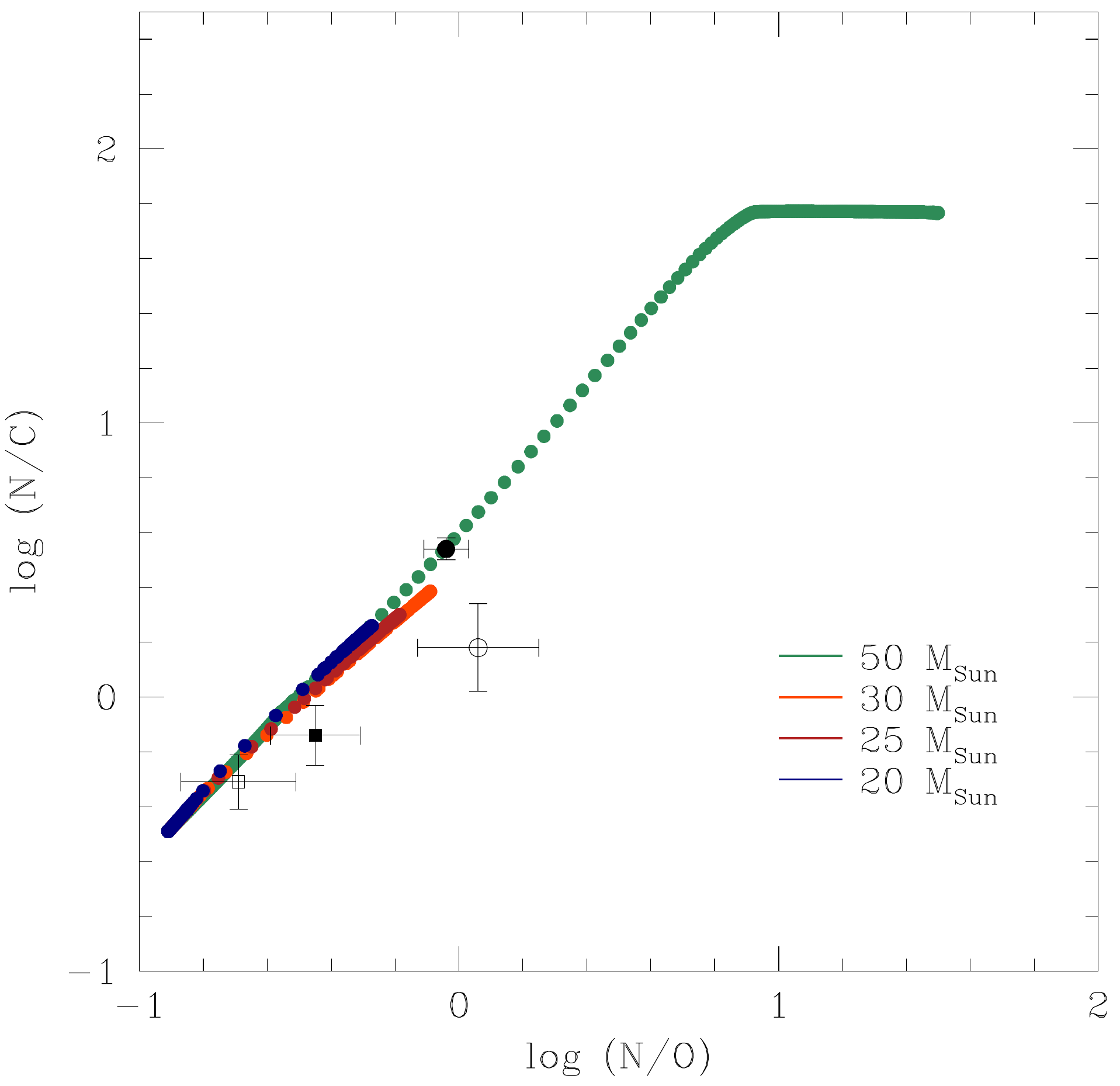}}
\end{minipage}
\caption{N/C and N/O ratios obtained from our spectral analysis. Squares and circles stand for the components of HD~17505Aa and HD~206267Aa, respectively. The filled (resp.\ empty) symbols correspond to the primary (resp.\ secondary) star. The coloured tracks show the predictions from solar metallicity single star evolutionary models of different initial masses. The left panel illustrates the comparison with models of \citet{Ekstrom12} without rotation, whereas the tracks in the middle panel correspond to the models of the same authors but assuming an initial rotational velocity of $0.4 \times v_{crit}$. The right panel shows the comparison with the models of \citet{Brott} accounting for rotation. In the right panel, initial rotational velocities are 111, 110, 109, and 213\,km\,s$^{-1}$ for the 20, 25, 30, and 50\,M$_{\odot}$ models, respectively.\label{figCNO}}
\end{figure*}

In Sect.\,\ref{Results-HD-17505} we found that the spectra of the components of HD 17505Aa display the signatures of a slight modification of the CNO surface abundances. Figure\,\ref{figCNO} compares our inferred N/C and N/O ratios with the predictions for evolutionary tracks of single massive stars from \citet{Ekstrom12} without rotation and with an initial rotational velocity of $0.4 \times v_{crit}$, as well as with models from \citet{Brott}. As demonstrated by \citet{Cazorla2}, these abundance ratios allow a more robust comparison with theoretical models than absolute abundances with respect to hydrogen. 

This figure shows that the current surface abundances of both stars are difficult to explain with models without rotation. Indeed, for such models, no modification of the CNO abundances is predicted on the main sequence for stars of initial mass below 60\,$M_{\odot}$. The spectroscopic masses of the components of HD~17505Aa are significantly lower than this number. On the other hand, the abundances of both stars can be perfectly well explained with single star models of the right initial mass when rotation is accounted for. We used the BONNSAI tool \citep{BONNSAI} to identify the single star rotating evolutionary tracks of \citet{Brott} that best match T$_{\rm eff}$, $\log{g}$, and $v\,\sin{i}$ of the stars investigated in this work. The results are provided in Table\,\ref{TabBonnsai}. As becomes clear from this table, these models fail however to predict the currently observed abundance ratios.

\begin{table}
\caption{BONNSAI results using T$_{\rm eff}$, $\log{g}$ and $v\,\sin{i}$ as input parameters and evolutionary tracks from \citet{Brott} \label{TabBonnsai}}
\begin{tabular}{c c c c c}
\hline
 & \multicolumn{2}{c}{HD~17505Aa} & \multicolumn{2}{c}{HD~206267Aa} \\
 & Prim. & Sec. & Prim. & Sec.\\
\hline
\vspace*{-2mm}\\
M$_{\rm init}$ (M$_{\odot}$) & $29.0^{+1.8}_{-1.5}$ & $30.2^{+1.8}_{1.6}$ & $42.0^{+7.8}_{-5.6}$ & $21.0^{+2.3}_{-2.0}$ \\
\vspace*{-2mm}\\
Age (Myr) & $3.2^{+0.6}_{-0.6}$ & $3.3^{+0.5}_{-0.5}$ & $2.0^{+0.5}_{-0.5}$ & $1.3^{+1.1}_{-1.1}$ \\
\vspace*{-2mm}\\
$v_{\rm rot, init}$ (km\,s$^{-1}$) & $70^{+58}_{-23}$ & $80^{+56}_{-28}$ & $200^{+114}_{-55}$ & $110^{+74}_{-19}$ \\
\vspace*{-2mm}\\
$\log{\rm (N/O)}$ & $-0.91^{+0.04}_{-0.04}$ & $-0.90^{+0.05}_{-0.05}$ & $-0.60^{+0.32}_{-0.32}$ & $-0.91^{+0.02}_{-0.02}$ \\
\vspace*{-2mm}\\
$\log{\rm (N/C)}$ & $-0.50^{+0.04}_{-0.04}$ & $-0.48^{+0.05}_{-0.05}$ & $-0.1^{+0.34}_{-0.34}$  & $-0.49^{+0.02}_{-0.02}$ \\
\vspace*{-2mm}\\
\hline
\end{tabular}
\end{table}

\begin{figure*}
\begin{minipage}{8cm}
\includegraphics*[width=\textwidth,angle=0]{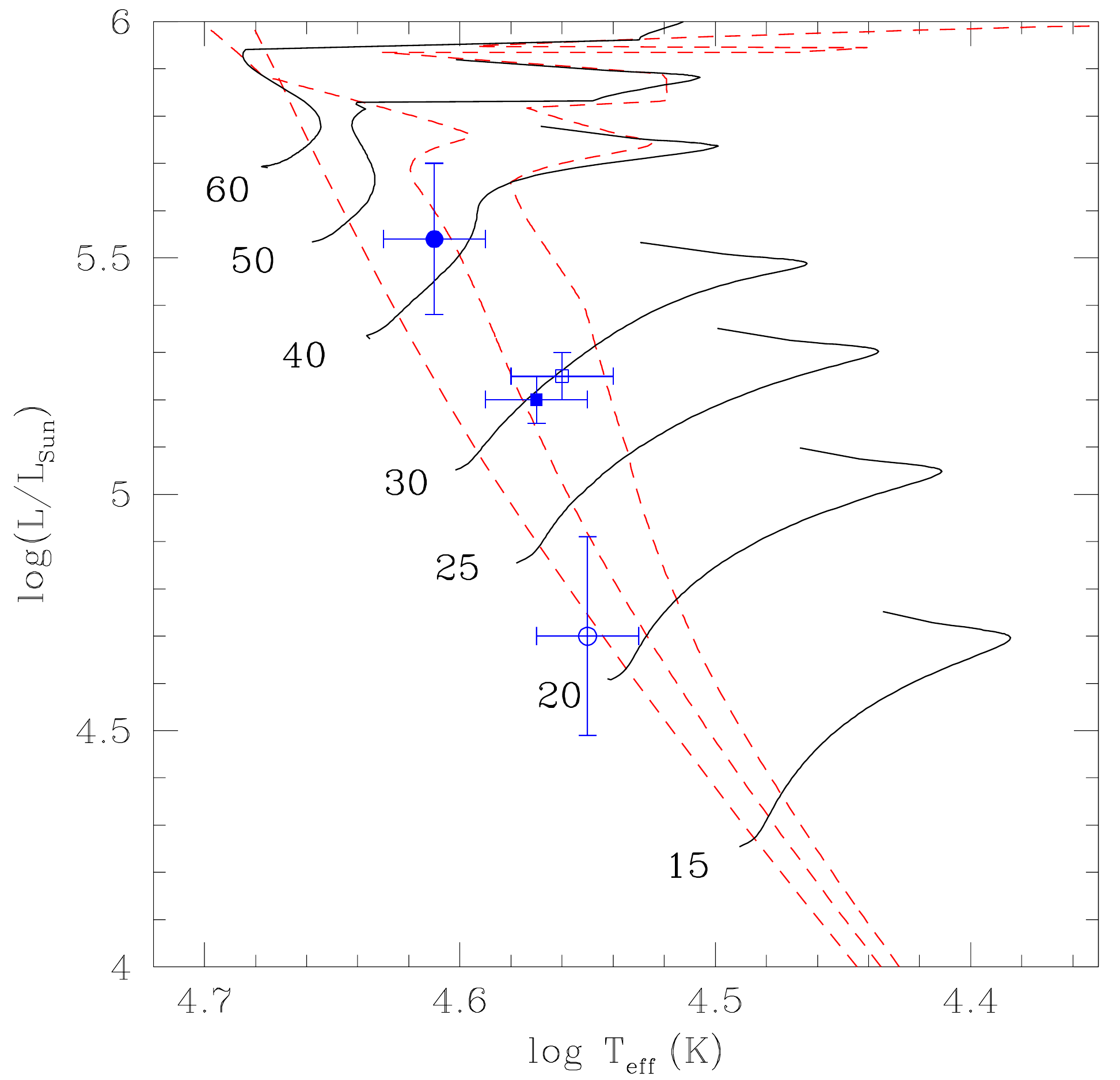}
\end{minipage}
\begin{minipage}{8cm}
\includegraphics*[width=\textwidth,angle=0]{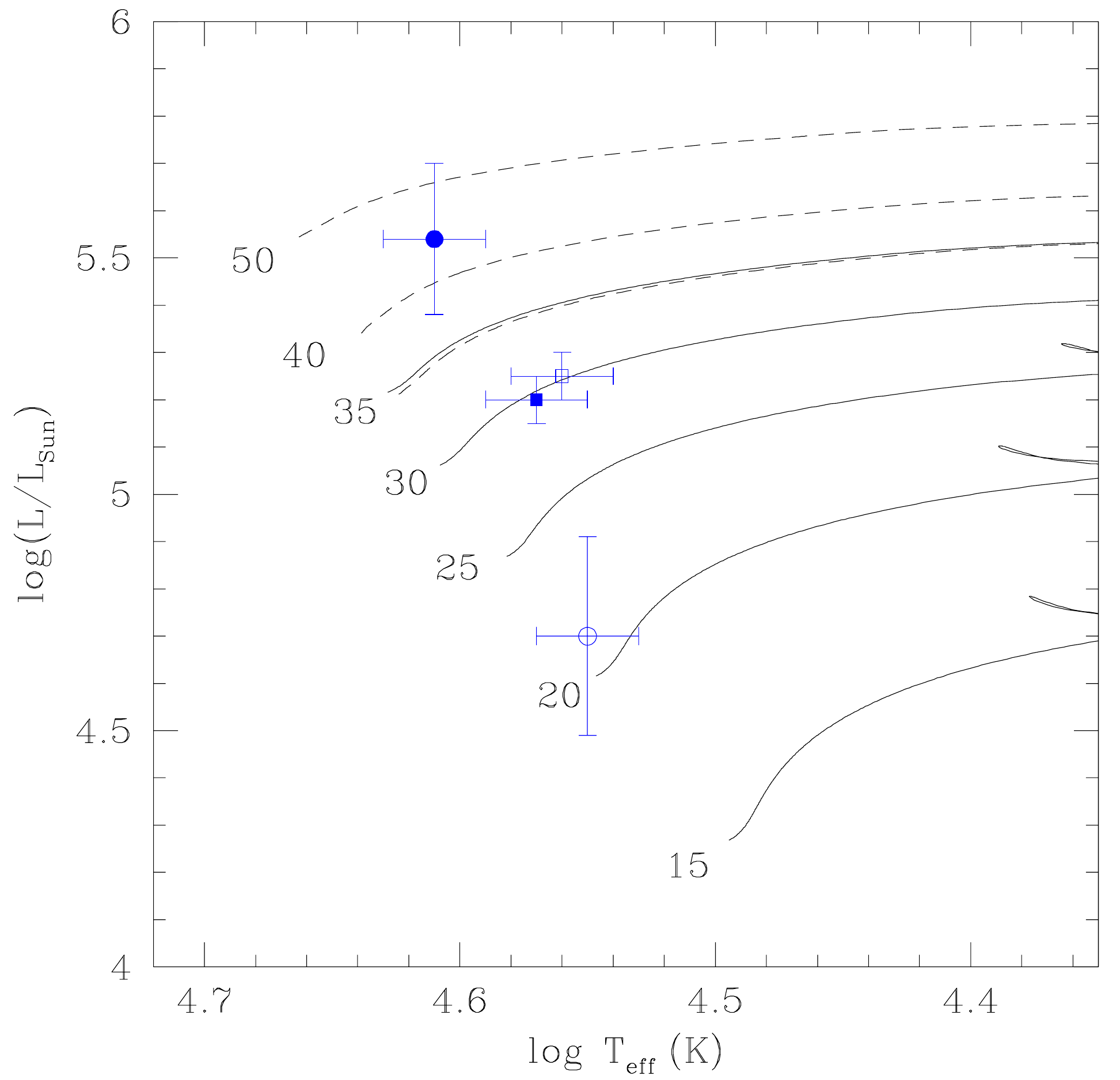}
\end{minipage}

\begin{minipage}{8cm}
\includegraphics*[width=\textwidth,angle=0]{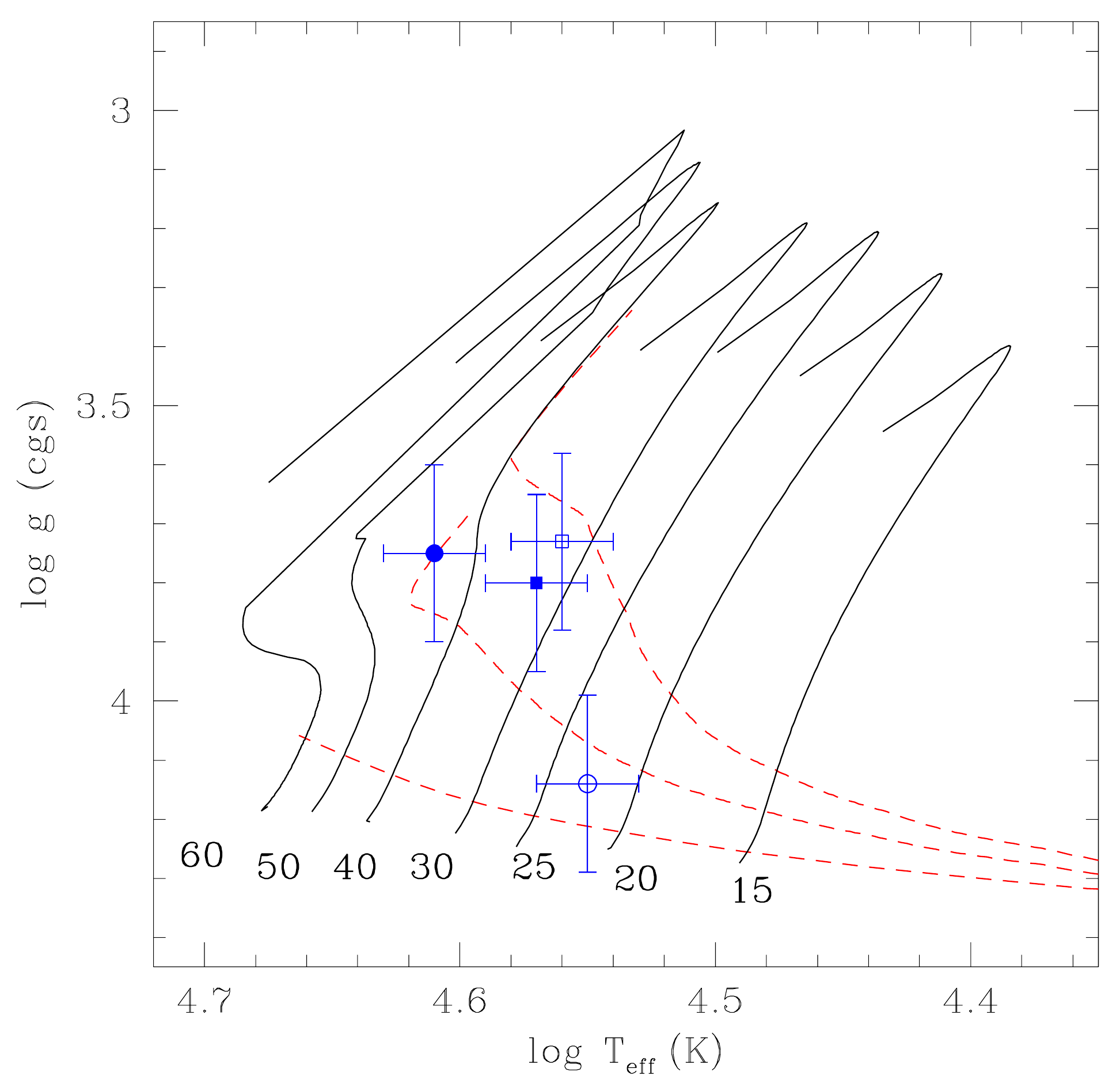}
\end{minipage}
\begin{minipage}{8cm}
\includegraphics*[width=\textwidth,angle=0]{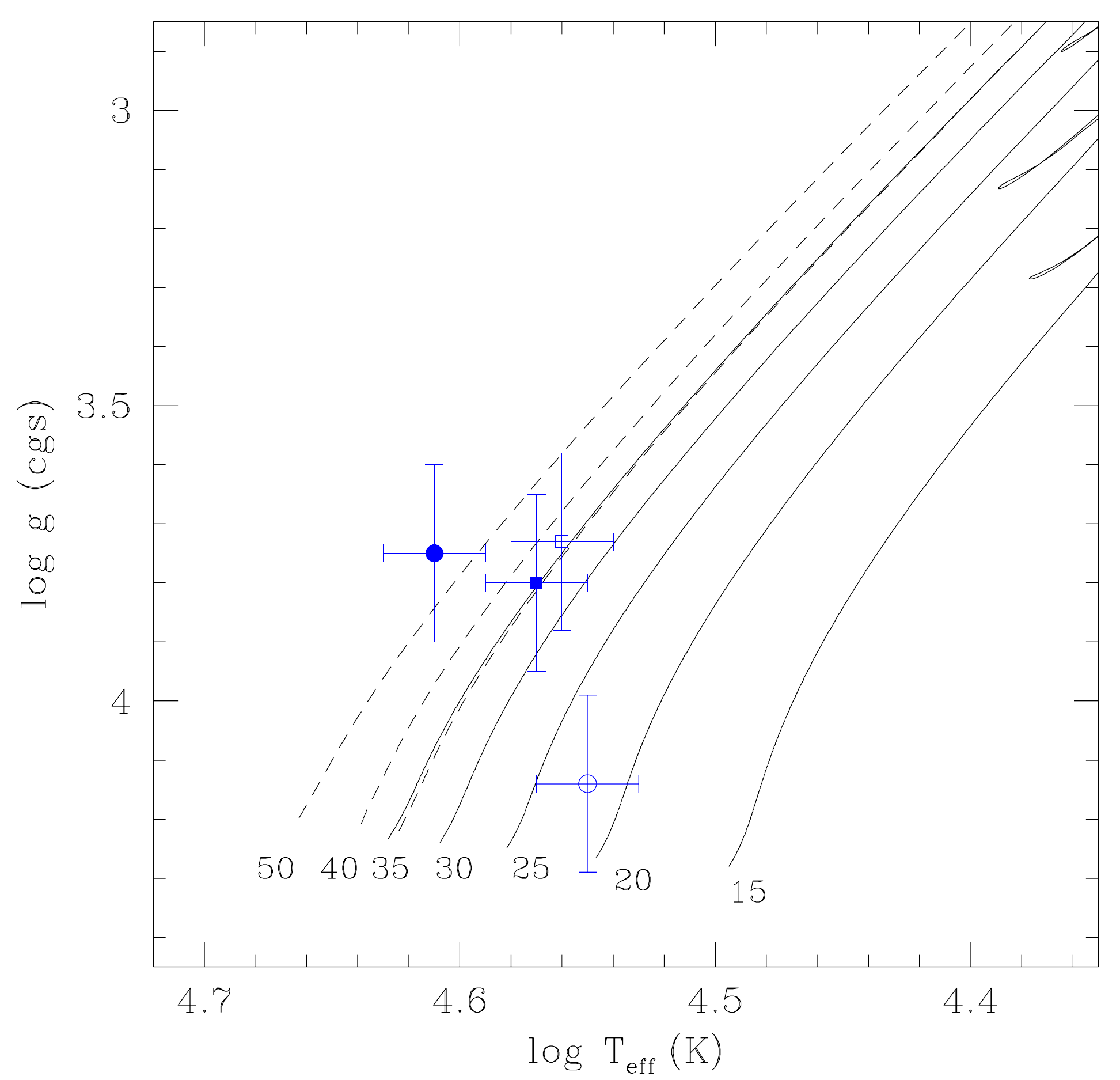}
\end{minipage}
\caption{Position of the stars in the Hertzsprung-Russell diagram (upper panels) and in the Kiel $\log{g}$ -- $\log{T_{\rm eff}}$ diagram (lower panels). Squares and circles stand for the components of HD~17505Aa and HD~206267Aa, respectively. The filled (resp.\ empty) symbols correspond to the primary (resp.\ secondary) star. In the left panels, evolutionary tracks for single massive stars at solar metallicity and initially rotating at $0.4 \times v_{crit}$ from \citet{Ekstrom12} are overplotted. The dashed red lines correspond to isochrones of 1.0, 3.2, and 5.0\,Myr. The right panels illustrate comparison with evolutionary tracks from \citet{Brott}. The solid lines yield tracks with initial rotational velocities near to 100\,km\,s$^{-1}$ (112, 111, 110, 109, and 109\,km\,s$^{-1}$ for the 15, 20, 25, 30, and 35\,M$_{\odot}$ models) whilst the dashed tracks correspond to initial rotational velocities near to 200\,km\,s$^{-1}$ (216, 214, and 213\,km\,s$^{-1}$ for the 35, 40, and 50\,M$_{\odot}$ models).
\label{figHRD}}
\end{figure*}

In Figure \ref{figHRD} we present the positions of the components of HD~17505Aa in the Hertzsprung-Russell diagram (HRD) and the $\log{g}$--$\log{T_{\rm eff}}$ Kiel diagram, along with the evolutionary tracks from \citet{Ekstrom12} with an initial rotational velocity of $0.4 \times v_{crit}$ and those of \citet{Brott} with initial rotational velocities that are nearest to those found in Table\,\ref{TabBonnsai}. 

The positions of the primary and secondary stars in the HRD suggest initial evolutionary masses near to 30\,$M_{\odot}$, which is larger than, but relatively close to, their current spectroscopic masses of 19--22\,$M_{\odot}$. The stars may be located on a common isochrone between 3.0 and 5.0\,Myr. 

Our results in Sect.\,\ref{Results-HD-17505} and the Figures\,\ref{figCNO} and \ref{figHRD} show that the current properties of the primary and secondary stars of HD7505Aa can most probably be explained by single star evolutionary models accounting for rotation. These considerations suggest that HD~17505Aa has not yet experienced a Roche lobe overflow (RLOF) binary interaction during its evolution.

\subsubsection{HD~206267Aa}
In Sect.\,\ref{Results-HD-206267}, we reported a modification of the CNO surface abundances of the primary and secondary components of HD~206267Aa. Comparison with single star models in Fig.\,\ref{figCNO} reveals that the abundances we determined are once again difficult to explain without rotational mixing, but are close to the predictions of rotating single star evolutionary models for initial masses close to 30 and 20 -- 25\,$M_{\odot}$ for the primary and secondary stars, respectively, which is reasonably close to the spectroscopic masses we have inferred (Table\,\ref{tableCMFGEN_206267}).

Their positions in the HRD and Kiel diagram (see Fig.\,\ref{figHRD}) suggest an initial mass slightly above 20\,$M_{\odot}$ for the secondary, which is well within the errors of its current spectroscopic mass of $17.7 \pm 8.9$\,$M_{\odot}$. The position of the primary star in the HRD and in the Kiel diagram suggests an initial mass between 40 and 50\,$M_{\odot}$, which is significantly higher than the spectroscopic mass of this star.  

Considering the slightly higher rotation rate of the secondary star of HD~206267Aa inferred in Sect.\,\ref{Results-HD-206267} and the fact that the primary is close to filling its Roche lobe around periastron passage as well as our analysis of Fig.\,\ref{figHRD}, the system may have encountered some kind of binary interactions during its evolution. However, the CNO abundances are only mildly altered; this suggests that the system did not yet experience a full case-A RLOF process, which should have affected the surface abundances of the donor star in a much stronger way \citep[see e.g.\ the cases of HD~149404 and LSS~3074,][]{Raucq16,Raucq17} and should also have led to orbital circularization, which is currently not the case. One way to explain the current status of HD~206267Aa would be an intermittent RLOF process around periastron, where mass transfer occurs temporarily, and is then interrupted until the next periastron passage. Tidal interactions in eccentric binaries actually lead to a more complex picture than what we can estimate based on the conventional Roche lobe model applied to a system with changing orbital separation \citep{Moreno}. Indeed, the tidal interactions around periastron can force oscillations that could help set up an intermittent transfer of matter. At this stage, we stress however that our existing spectra of HD~206267 do not reveal any obvious observational signature of such an interaction, such as H$\alpha$ emission, which would appear at phases near to periastron. Therefore, if such intermittent RLOF has taken place, it might have ceased now, although a more dense spectroscopic monitoring of the phases around periastron is probably required to unveil a very short-lived periastron event.

One possible reason for the interruption of such a process could be the dynamical influence of the third component in a hierarchical triple system. Indeed, the presence of a third component in the system can lead to the appearance of Lidov-Kozai cycles \cite[][and references therein]{Toonen16}. In such cases, angular momentum exchange between the inner and outer orbits occurs owing to a mutual torque between these orbits. Since the orbital energy, and therefore the semi-major axes, are conserved, the orbital eccentricity of the inner binary system and the mutual inclination between the two orbits can vary periodically. This Lidov-Kozai mechanism could therefore lead to a periodic modulation of the binary interaction at periastron passage.

Given the angular separation of 0.1\,arcsec at an assumed distance of 1050\,pc, we find a separation of at least 105\,AU between the inner binary system and the third component. The orbital period of the third component would thus be at least 135\,years. Since the timescale of the Lidov-Kozai cycle scales with the square of the ratio of the outer to inner orbital periods \citep{Toonen16}, we estimate a minimum timescale of the order 1.75\,Myr. The fact that this timescale is very long and the wide separation between the inner binary and the third component make a strong influence of the tertiary on the evolution of the close binary rather unlikely. 

\subsection{Summary and conclusions}
We have studied the fundamental properties of the inner binary systems of the triple systems HD~17505A and HD~206267A. We first improved the orbital solutions of the inner binary systems. We then fitted the spectra of the triple systems with a combination of synthetic CMFGEN spectra of the three components shifted by their associated RVs for each observation to subtract the third component from the observed spectra and recover the spectra of the inner binary. We subsequently used our disentangling code to recover the individual spectra of the primary and secondary stars for both systems. From these reconstructed spectra, we determined a number of stellar parameters, partly with the CMFGEN model atmosphere code, and used them to constrain the evolutionary status of the systems.

We found that the CNO abundances and the properties of the primary and secondary spectra of HD~17505Aa can be explained by single star evolutionary models of initial mass of about 30\,$M_{\odot}$ accounting for rotational mixing. At this stage of its evolution, this system has thus not yet experienced binary interactions.

Whilst the CNO abundances of the components of HD~206267Aa can be similarly explained by single star evolutionary models accounting for rotational mixing, we found that the secondary star of this system displays a slightly higher rotation rate. Furthermore, the primary appears slightly overluminous for its mass. This suggests that the system could have experienced intermittent binary interactions around periastron in the past, but has not yet experienced a complete RLOF episode. In future studies, it would be highly interesting to collect photometric data of HD~206267A. Indeed, with a period near to 3.7\,days and an estimated inclination of $76^{\circ}$, it appears likely that the inner binary system displays ellipsoidal variations that might help to further constrain its parameters. Moreover, establishing whether or not the third component seen in the spectrum is gravitationally bound to the inner binary and deriving its orbital parameters would be important to check whether or not Lidov-Kozai cycles played a role in shaping the properties of the system as we observe it today.

\acknowledgement{We acknowledge financial support through an ARC grant for Concerted Research Actions, financed by the Federation Wallonia-Brussels, from the Fonds de la Recherche Scientifique (FRS/FNRS) through an FRS/FNRS Research Project (T.0100.15), and through an XMM PRODEX contract (Belspo). We are grateful to Pr.\ D.J.\ Hillier for making the CMFGEN model atmosphere code available. Computational resources have been provided by the Consortium des Equipements de Calcul Intensif (C\'ECI), funded by the FRS/FNRS under Grant No.\ 2.5020.11. We thank the anonymous referee for a careful and constructive report.}

\end{document}